\documentclass[prd,tightenlines,nofootinbib,superscriptaddress]{revtex4}

\usepackage{amsfonts,amssymb,amsthm,bbm,amsmath}
\usepackage{hyperref}

\usepackage{color,psfrag}
\usepackage[dvips]{graphicx}

\usepackage{tikz}

\usepackage{subcaption}
\newcommand{\C}{{\mathbb C}}
\newcommand{\N}{{\mathbb N}}
\newcommand{\R}{{\mathbb R}}

\newcommand{\cV}{{\mathcal V}}

\newcommand{\cC}{{\mathcal C}}
\newcommand{\cS}{{\mathcal S}}
\newcommand{\cU}{{\mathcal U}}
\newcommand{\cZ}{{\mathcal Z}}

\newcommand{\SU}{\mathrm{SU}}

\newcommand{\U}{\mathrm{U}}

\newcommand{\be}{\begin{equation}}
\newcommand{\ee}{\end{equation}}
\newcommand{\beq}{\begin{eqnarray}}
\newcommand{\eeq}{\end{eqnarray}}
\newcommand{\bes}{\begin{eqnarray}}
\newcommand{\ees}{\end{eqnarray}}

\newcommand{\sixj} [6] {\left\{ \begin{array}{ccc}{#1}&{#2}&{#3}\\{#4}&{#5}&{#6}\end{array} \right\} }

\newcommand{\ra}{\rangle}

\newcommand{\tr}{{\mathrm{Tr}}}
\newcommand{\f}{\frac}

\def\nn{\nonumber}
\def\pp{\partial}
\newcommand{\w}{\wedge}

\def\vphi{\varphi}
\def\eps{\epsilon}

\newcommand{\id}{\mathbb{I}}

\def\vJ{\vec{J}}

\def\rd{\mathrm{d}}
\def\vv{\vec{v}}
\def\vw{\vec{w}}

\begin{document}

\title{3d Quantum Gravity: Coarse-Graining and $q$-Deformation}

\author{{\bf Etera R. Livine}}\email{etera.livine@ens-lyon.fr}
\affiliation{Univ Lyon, Ens de Lyon, Universit\'e Claude Bernard Lyon 1, CNRS,
Laboratoire de Physique, F-69342 Lyon, France}
\affiliation{Perimeter Institute for Theoretical Physics, 31 Caroline Street North, Waterloo, Ontario, Canada N2L 2Y5}

\date{\today}

\begin{abstract}

The Ponzano-Regge state-sum model provides a quantization of 3d gravity as a spin foam, providing a quantum amplitude to each 3d triangulation defined in terms of the 6j-symbol (from the spin-recoupling theory of $\SU(2)$ representations).
In this context, the invariance of the 6j-symbol under 4-1 Pachner moves, mathematically defined by the Biedenharn-Elliot identity, can be understood as the invariance of the Ponzano-Regge model under coarse-graining or equivalently as the invariance of the amplitudes under the Hamiltonian constraints.
Here we look at length and volume insertions in the Biedenharn-Elliot identity for the 6j-symbol, derived in some sense as higher derivatives of the original formula.  This gives the behavior of these geometrical observables under coarse-graining.
These new identities turn out to be related to the Biedenharn-Elliot identity for the $q$-deformed 6j-symbol and highlight that the $q$-deformation produces a cosmological constant term in the Hamiltonian constraints of 3d quantum gravity.

\end{abstract}

\maketitle


The Ponzano-Regge model \cite{Freidel:2004vi,Freidel:2004nb,Freidel:2005bb,Barrett:2006ru,Barrett:2008wh} realizes the quantization of 3d gravity for a Euclidean space-time signature. It proposes a discrete path integral understood as the spinfoam quantization of 3d gravity written in its first order formulation as a BF theory (for reviews of the spinfoam framework, see e.g. \cite{Livine:2010zx,Alexandrov:2011ab,Perez:2012wv}).  It assigns an amplitude to every three-dimensional triangulation by
associating to every tetrahedron a 6j-symbol from the recoupling theory of spins, i.e. of $\SU(2)$ representations. That amplitude is understood to be topologically invariant and only depend on the topology of the 3d triangulation and the boundary data.
When the triangulation represents the evolution of a spatial slice, i.e. is topologically equivalent to a cylinder $\Sigma_{\textrm{2d}}\times [0,1]$, the Ponzano-Regge model is shown to implement the Hamiltonian constraints -the dynamics- of 3d loop quantum gravity. Indeed, in that setting, the amplitude defines a projector on the moduli space of flat $\SU(2)$ connections on the spatial slice $\Sigma$, which constitute the space of physical states of the theory. 
Finally, the large spin asymptotics of the 6j-symbol (see e.g. \cite{Schulten:1971yv,Schulten:1975yu,Freidel:2002mj,Roberts:2002,Gurau:2008yh}) allows to interpret the Ponzano-Regge amplitude as a quantized version of a path integral for Regge calculus. 

It is possible to $q$-deform the Ponzano-Regge model by using the representation of the quantum group $\cU_{q}\SU(2)$ or $\SU_{q}(2)$. This leads to the Turaev-Viro model \cite{Turaev:1992hq}, based on the $q$-deformed $\{6j\}_{q}$-symbol (see also \cite{Dittrich:2016typ} for a recent description of the Turaev-Viro amplitudes and combinatorics, together with a detailed analysis of particle defects). The asymptotics of the $\{6j\}_{q}$-symbol proposes the interpretation of the $q$-deformation as a non-vanishing cosmological constant. Recent work on the canonical interpretation of the theory from the viewpoint of loop quantum gravity, studying how to map a homogeneous $\SU(2)$ curvature to flat $\SU_{q}(2)$ connections, tend to confirm this claim for both a positive cosmological $\Lambda>0$ corresponding to $q$ root of unity \cite{Noui:2011im,Noui:2011aa,Pranzetti:2014xva} and for a negative cosmological constant $\Lambda<0$ corresponding to a real deformation parameter $q$  \cite{Dupuis:2013haa,Dupuis:2013lka,Bonzom:2014bua,Bonzom:2014wva}.

Here we study further this relation between the $q$-deformation and the cosmological constant in 3d gravity. We build on the previous work \cite{Freidel:1998ua} by Freidel and Krasnov, looking at the deformation of the 6j-symbol into  $\{6j\}_{q}$-symbol at leading order in $q$ and relating it to the volume operator acting on the 6j-symbol, and we re-visit this relation in the new light of the invariance under coarse-graining of the Ponzano-Regge model.

\medskip

In this paper, we focus on the invariance of the Ponzano-Regge and Turaev-Viro models under 4-1 Pachner moves, which is a particular case of the topological invariance. A 4-1 Pachner move splits a single tetrahedron into four tetrahedra by introducing a new bulk vertex, as illustrated on fig.\ref{fig41}, and can be geometrically interpreted as the basic coarse-graining or refining move for a 3d triangulation. The invariance of the 
Ponzano-Regge amplitude under 4-1 Pachner move mathematically materializes in the Biedenharn-Elliott identity satisfied by the 6j-symbol. 
We first review how this 4-1 Pachner move invariance can be derived, in the context of loop quantum gravity, from the action of the holonomy operator on  tetrahedral spin networks evaluated on the flat state as shown in \cite{Bonzom:2009zd}.  This is physically-relevant since the holonomy operators are in fact the Hamiltonian constraints for 3d quantum gravity \cite{Noui:2004iy,Noui:2004ja}. 

Then the heart of the paper is the study of the behavior of lengths and volumes under coarse-graning, defined as the propagation of these geometrical observables through 4-1 Pachner moves. We analyze the action of length and volume operators on the 6j-symbol, following the idea of \cite{Charles:2016xwc} of considering higher derivatives of the flat state. This leads to new extended Biedenharn-Elliott identities with length and volume insertions satisfied by the 6j-symbol.

Finally, we show how these new formulas are actually related to the 4-1 Pachner move invariance of the $\{6j\}_{q}$-symbol.
More precisely, we look at the first order of expansion in the deformation parameter $q$ and explain how the variation between the 6j-symbol and the $\{6j\}_{q}$-symbol is given at leading order by the volume operator acting on 6j-symbol. We perform numerical tests of this first order correction formula using the explicit expressions for the  triple graspings of the 6j-symbols given in \cite{Hackett:2006gp}.
This provides a clear geometrical interpretation of $\{6j\}_{q}$-symbols and further confirms the interpretation of the $q$-deformation as accounting for a cosmological constant in 3d quantum gravity. 
And it allows us to prove the 4-1 Pachner move invariance of the $\{6j\}_{q}$-symbol from the coarse-graining formulas for  lengths and volumes. This provides us with a very simple geometrical interpretation of the $q$-deformed Biedenharn-Elliott identity as the quantum expression of the conservation of volume under a 4-1 Pachner move.

\section{4$\leftrightarrow$1 Pachner move and Coarse-Graining Observables}

\begin{figure}[h!]
\centering
\begin{tikzpicture}[scale=1.5]
\coordinate (O1) at (0,0,0);

\coordinate (A1) at (0,1.061,0);
\coordinate (B1) at (0,-0.354,1);
\coordinate (C1) at (-0.866,-0.354,-0.5);
\coordinate (D1) at (0.866,-0.354,-0.5);

\draw[blue] (A1) -- (B1) node[midway,right]{$j_{6}$};
\draw[blue] (A1) -- (C1)node[midway,left]{$j_{5}$};
\draw[blue] (A1) -- (D1)node[midway,right]{$j_{4}$};
\draw[blue] (B1) -- (C1)node[midway,left]{$j_{1}$};
\draw[dashed,blue] (C1) -- (D1)node[midway,above]{$j_{3}$};
\draw[blue] (D1) -- (B1)node[midway,below]{$j_{2}$};


\draw[blue] (A1) node{$\bullet$};
\draw[blue] (B1) node{$\bullet$};
\draw[blue] (C1) node{$\bullet$};
\draw[blue] (D1) node{$\bullet$};

\end{tikzpicture}

\caption{A tetrahedron with spins living on its six edges.}
\label{fig_tetra}
\end{figure}
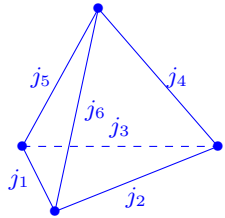

The Ponzano-Regge model associates an amplitude to every three-dimensional triangulation\footnotemark{} with a boundary.
\footnotetext{It is possible to extend the definition of the Ponzano-Regge model to arbitrary 3d cellular complexes. This is best formalized within the spinfoam framework (see \cite{Livine:2010zx,Alexandrov:2011ab,Perez:2012wv} for reviews). }
A 3d triangulation is defined as a set of tetrahedra glued together along their triangles. We attach a half-integer spin $j_{e}\in\N/2$ to each edge of the triangulation. Each spin corresponds to a unitary irreducible representation of $\SU(2)$. This spin is understood geometrically as the quantization of the length edge in Planck units:
\be
\ell_{e}= j_{e} l_{P}\,.
\ee
We write $\cV^j$ for the Hilbert space carrying the spin-$j$ representation. It has dimension $d_{j}=(2j+1)$ and $\SU(2)$ Casimir $\cC_{j}=j(j+1)$. Considering a tetrahedron, it is made of six edges, each carrying a spin $j_{i=1..6}$, as illustrated on fig.\ref{fig_tetra}. We associate to the tetrahedron the Wigner 6j-symbol. It is defined as the contraction of four Clebsch-Gordan coefficients,  corresponding to the four triangles of the tetrahedron:
\be
\sixj{j_{1}}{j_{2}}{j_{3}}{j_{4}}{j_{5}}{j_{6}}
=\,
\sum_{\{m_{i}\}}
(-1)^{\sum_{i=1}^{6}j_{i}-m_{i}}\,
C^{j_{1}j_{2}j_{3}}_{m_{1},m_{2},-m_{3}}
C^{j_{1}j_{5}j_{6}}_{-m_{1},m_{5},m_{6}}
C^{j_{4}j_{5}j_{3}}_{m_{4},-m_{5},m_{3}}
C^{j_{4}j_{2}j_{6}}_{-m_{4},-m_{2},-m_{6}}\,,
\ee
where the $C$'s are the Wigner 3j-symbol, defining the unique 3-valent intertwiners between spin triplets (see in appendix \ref{def3j} for more details) and obtained from the Clebsch-Gordan coefficients from a simple renormalization.
This 6j-symbol can be given a more straightforward expression as a sum over a single integer, although it is usually more convenient to compute it using its recursion relation \cite{Schulten:1975yu} when doing numerical simulations. This is Racah's sum formula\footnotemark{}:
\be
\sixj{j_{1}}{j_{2}}{j_{3}}{j_{4}}{j_{5}}{j_{6}}
=\,
\sqrt{\Delta_{j_{1}j_{2}j_{3}}\Delta_{j_{1}j_{5}j_{6}}\Delta_{j_{4}j_{2}j_{6}}\Delta_{j_{4}j_{5}j_{3}}}\,
\sum_{t=\max{\alpha_{k}}}^{\min{\beta_{l}}}
(-1)^{t}
\f{(t+1)!}{\prod_{k=1}^{4}(t-\alpha_{k})!\,\prod_{l=1}^{3}(\beta_{l}-t)!}
\,,
\ee
where the $\alpha$'s and $\beta$'s are respectively the sum of spins around the triangles and 4-cycles of the tetrahedron:
$$
\alpha_{k}=
\left|
\begin{array}{l}
j_{1}+j_{2}+j_{3}\\
j_{1}+j_{5}+j_{6}\\
j_{4}+j_{2}+j_{6}\\
j_{4}+j_{5}+j_{3}
\end{array}
\right.
\,,\qquad
\beta_{l}=
\left|
\begin{array}{l}
j_{1}+j_{2}+j_{4}+j_{5}\\
j_{1}+j_{3}+j_{4}+j_{6}\\
j_{2}+j_{3}+j_{5}+j_{6}
\end{array}
\right.\,,
$$
and the $\Delta_{abc}$ are the standard triangular coefficients, attached to each triangle of the tetrahedron, defined as:
$$
\Delta_{abc}=
\f{(-a+b+c)!\,(a-b+c)!\,(a+b-c)!}{(a+b+c+1)!}
\,.
$$
\footnotetext{In \cite{Gurau:2008yh}, the Regge asymptotics of the 6j-symbol for large spins was derived studying the saddle point approximation of the the Racah sum formula and expressing the saddle points in terms of the edge lengths of the tetrahedron determining its geometry.}
The Ponzano-Regge amplitude of a triangulation $\Delta$ is then defined as the sum over all bulk spins of the product of the 6j-symbols corresponding to the tetrahedra (see \cite{Barrett:2008wh} for details and explanations on the sign factors, also see \cite{Livine:2003hn} for their interpretation in terms of supersymmetry):
\be
\cZ^{PR}[\Delta]=\sum_{\{j_{e}\}} \prod_{e}(-1)^{2j_{e}}(2j_{e}+1)\,\prod_{T}
(-1)^{\sum_{i=1}^{6}j_{i}^{T}}\sixj{j_{1}^{T}}{j_{2}^{T}}{j_{3}^{T}}{j_{4}^{T}}{j_{5}^{T}}{j_{6}^{T}}\,.
\label{PR}
\ee
This amplitude is also derived as a discrete path integral for 3d gravity, yielding its spinfoam quantization, and leads back to Regge calculus for discrete gravity in its large spin regime, due to the celebrated asymptotic formula for the 6j symbol (see e.g. \cite{Freidel:2002mj}):
\be
\{6j\}\sim\,
\f1{\sqrt{12\pi \,V}}\,\cos\left(S_{\textrm{Regge}}+\f\pi4\right)\,,
\qquad\textrm{with}\quad S_{\textrm{Regge}}=\sum_{i}\left(j_{i}+\f12\right)\,\theta_{i}\,,
\ee
where  $\theta_{i}$ is the (external) dihedral angle associated to the edge $i$ (the angle between the planes of the two triangles sharing that edge) as a function of the edge lengths of the tetrahedron identified as $\ell_{i}=j_{i}+\f12$. The volume $V$ is also the volume of that tetrahedron. The shift $+\f12$ between the spins $j_{i}$ and the classical edge lengths $\ell_{i}$ is a semi-classical correction, which can be checked to significantly improve any numerical simulation of the 6j-symbol at large spins.

\medskip

The Ponzano-Regge amplitude \eqref{PR} is topologically invariant and also typically infinite. These two features are intimately intertwined and reflect the situation of the continuum theory. Indeed, it is understood that the divergence of the Ponzano-Regge model is due to the non-compact translational invariance (Poincar\'e symmetry) of 3d gravity, which leads to the topological invariance of 3d quantum gravity. As shown in \cite{Freidel:2004vi,Barrett:2008wh}, it is possible to gauge-fix this symmetry and define meaningful finite amplitudes (for more detail and thorough analysis, we refer the interested reader to \cite{Bonzom:2010ar,Bonzom:2010zh,Bonzom:2012mb}). This gauge-fixing (with a trivial Fadeev-Popov determinant) amounts to fixing the spins $j_{e}$ on a certain number of edges. When that spin is fixed to 0, this is actually equivalent of contracting the edge, thus locally collapsing the triangulation. More precisely, the divergences are related to the presence of bulk vertices in the 3d triangulation \cite{Freidel:2004vi}. It is possible to collapse the triangulation and effectively remove all bulk vertices, thus leading to a convergent Ponzano-Regge amplitude.

\smallskip

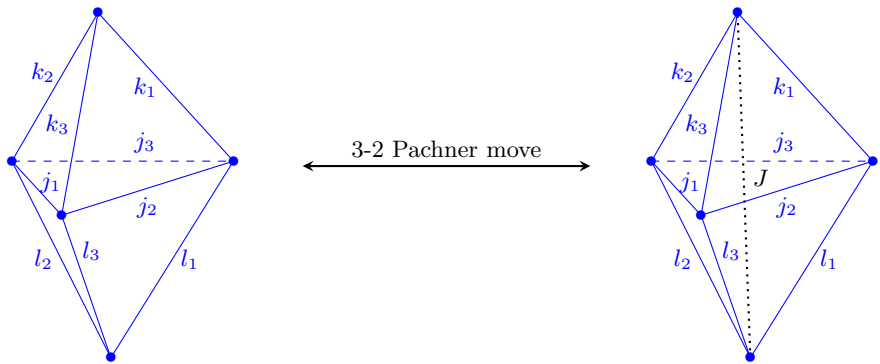
\begin{figure}[h!]
\centering
\begin{tikzpicture}[scale=1.7]

\coordinate (A1) at (0,1,0);
\coordinate (B1) at (0.1,-0.2,1);
\coordinate (C1) at (-0.866,-0.354,-0.5);
\coordinate (D1) at (0.866,-0.354,-0.5);
\coordinate (E1) at (0.1,-1.7,0);

\draw[blue] (A1) -- (B1) node[pos=0.55,left]{$k_{3}$};
\draw[blue] (A1) -- (C1) node[pos=0.4,left]{$k_{2}$};
\draw[blue] (A1) -- (D1) node[midway,left]{$k_{1}$};
\draw[blue] (B1) -- (C1) node[pos=0.6,right]{$j_{1}$};
\draw[dashed,blue] (C1) -- (D1)node[pos=0.6,above]{$j_{3}$};
\draw[blue] (D1) -- (B1)node[midway,below]{$j_{2}$};

\draw[blue] (E1) -- (B1) node[near end,right]{$l_{3}$};
\draw[blue] (E1) -- (C1) node[midway,left]{$l_{2}$};
\draw[blue] (E1) -- (D1) node[midway,right]{$l_{1}$};

\draw[blue] (A1) node{$\bullet$};
\draw[blue] (B1) node{$\bullet$};
\draw[blue] (C1) node{$\bullet$};
\draw[blue] (D1) node{$\bullet$};
\draw[blue] (E1) node{$\bullet$};

\coordinate (A2) at (5,1,0);
\coordinate (B2) at (5.1,-0.2,1);
\coordinate (C2) at (4.134,-0.354,-0.5);
\coordinate (D2) at (5.866,-0.354,-0.5);
\coordinate (E2) at (5.1,-1.7,0);

\draw[blue] (A2) -- (B2)node[pos=0.55,left]{$k_{3}$};
\draw[blue] (A2) -- (C2) node[pos=0.4,left]{$k_{2}$};
\draw[blue] (A2) -- (D2)node[midway,left]{$k_{1}$};
\draw[blue] (B2) -- (C2)node[pos=0.6,right]{$j_{1}$};
\draw[dashed,blue] (C2) -- (D2)node[pos=0.6,above]{$j_{3}$};
\draw[blue] (D2) -- (B2)node[midway,below]{$j_{2}$};

\draw[blue] (E2) -- (B2) node[near end,right]{$l_{3}$};
\draw[blue] (E2) -- (C2)node[midway,left]{$l_{2}$};
\draw[blue] (E2) -- (D2)node[midway,right]{$l_{1}$};

\draw[dotted,thick] (E2) -- (A2) node[pos=0.52,right]{$J$};

\draw[blue] (A2) node{$\bullet$};
\draw[blue] (B2) node{$\bullet$};
\draw[blue] (C2) node{$\bullet$};
\draw[blue] (D2) node{$\bullet$};
\draw[blue] (E2) node{$\bullet$};

\draw[<->,>=stealth, thick] (1.6,-0.2,0) to node[midway,above]{3-2 Pachner move} (3.85,-0.2,0);

\end{tikzpicture}

\caption{3-2 Pachner move}
\label{fig32}
\end{figure}
All this is best illustrated by the invariance of the Ponzano-Regge amplitude under 3-2 and 4-1 Pachner moves, which are the mathematical identities expressing the topological invariance of the model. These moves allow to relate any two topologically equivalent triangulations through a finite sequence of moves.
The 3-2 Pachner move, as illustrated on fig.\ref{fig32}, replaces three tetrahedra around one bulk edge with two tetrahedron. The invariance of the Ponzano-Regge model is expressed by the Biedenharn-Elilot identity, also known as the pentagonal identity:
\be
\sixj{j_{1}}{j_{2}}{j_{3}}{k_{1}}{k_{2}}{k_{3}}\sixj{j_{1}}{j_{2}}{j_{3}}{l_{1}}{l_{2}}{l_{3}}
\,=\,
\sum_{J}
(-1)^{\Phi}\,(2J+1)\,
\sixj{j_{1}}{l_{2}}{l_{3}}{J}{k_{3}}{k_{2}}
\sixj{l_{1}}{j_{2}}{l_{3}}{k_{3}}{J}{k_{1}}
\sixj{l_{1}}{l_{2}}{j_{3}}{k_{2}}{k_{1}}{J}
\,,
\label{32move} 
\ee
where where we sum over the spin $J$ on the bulk edge and the sign factor is given by:
$$
\Phi=j_{1}+j_{2}+j_{3}+k_{1}+k_{2}+k_{3}+l_{1}+l_{2}+l_{3}+J\,.
$$
Using this identity, one can indeed check explicitly the invariance of the partition function \eqref{PR} under the 3-2 Pachner move.

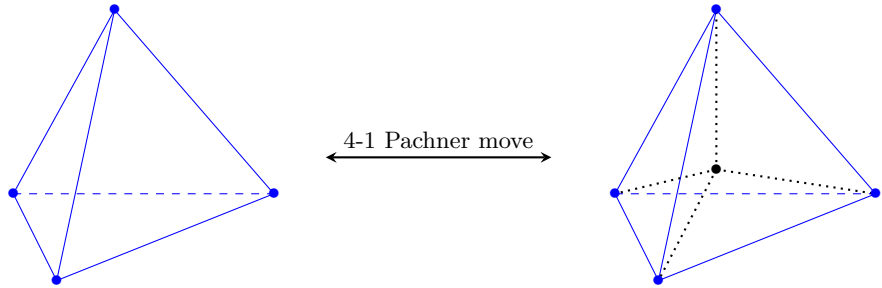
\begin{figure}[h!]
\centering
\begin{tikzpicture}[scale=2]

\coordinate (A1) at (0,1.061,0);
\coordinate (B1) at (0,-0.354,1);
\coordinate (C1) at (-0.866,-0.354,-0.5);
\coordinate (D1) at (0.866,-0.354,-0.5);

\draw[blue] (A1) -- (B1);
\draw[blue] (A1) -- (C1);
\draw[blue] (A1) -- (D1);
\draw[blue] (B1) -- (C1);
\draw[dashed,blue] (C1) -- (D1);
\draw[blue] (D1) -- (B1);

\draw[blue] (A1) node{$\bullet$};
\draw[blue] (B1) node{$\bullet$};
\draw[blue] (C1) node{$\bullet$};
\draw[blue] (D1) node{$\bullet$};

\coordinate (O2) at (4,0,0);

\coordinate (A2) at (4,1.061,0);
\coordinate (B2) at (4,-0.354,1);
\coordinate (C2) at (3.134,-0.354,-0.5);
\coordinate (D2) at (4.866,-0.354,-0.5);

\draw[blue] (A2) -- (B2);
\draw[blue] (A2) -- (C2);
\draw[blue] (A2) -- (D2);
\draw[blue] (B2) -- (C2);
\draw[dashed,blue] (C2) -- (D2);
\draw[blue] (D2) -- (B2);

\draw[dotted, thick] (O2) -- (A2);
\draw[dotted, thick] (O2) -- (B2);
\draw[dotted, thick] (O2) -- (C2);
\draw[dotted, thick] (O2) -- (D2);

\draw (O2) node{$\bullet$};
\draw[blue] (A2) node{$\bullet$};
\draw[blue] (B2) node{$\bullet$};
\draw[blue] (C2) node{$\bullet$};
\draw[blue] (D2) node{$\bullet$};

\draw[<->,>=stealth, thick] (1.5,0.177,0.25) to node[midway,above]{4-1 Pachner move} (3,0.177,0.25);

\end{tikzpicture}

\caption{4-1 Pachner move}
\label{fig41}
\end{figure}
The 4-1 Pachner move, as  illustrated on fig.\ref{fig41}, replaces a single tetrahedron with four tetrahedra by adding one bulk vertex. Adding and removing bulk vertices is the fundamental method of coarse-graining the Ponzano-Regge amplitude.
Summing over the spins living on the four bulk edges is actually divergent. This is the basic reason of the divergence of the Ponzano-Regge amplitude. This is due to presence of the bulk vertex. It is nevertheless possible to make this move finite by fixing one of the four internal spins in the sum. As shown in \cite{Bonzom:2009zd} and reviewed in detail below, this finite 4-1 Pachner move identity can be derived from the action of  the holonomy operator on the 6j-symbol and shows the invariance of the Ponzano-Regge path integral under coarse-graining.

\subsection{The 4$\leftrightarrow$1 Pachner move from the Holonomy Operator}

To formalize the 4-1 Pachner move, we start with the spin network state living on the tetrahedron graph. Labeled by the six spins $j_{i}$, it is defined as a function of six group elements $g_{i}\in \SU(2)$ living on the dual links:
\be
\vphi_{\{j_{i}\}}(\{g_{i}\})
\,=\,
\sum_{\{m_{i},n_{i}\}}
\prod_{i}^{6} D^{j_{i}}_{m_{i}n_{i}}(g_{i})
\,
(-1)^{\sum_{i=1}^{6}j_{i}-n_{i}}\,
C^{j_{1}j_{2}j_{3}}_{m_{1},m_{2},-n_{3}}
C^{j_{1}j_{5}j_{6}}_{-n_{1},m_{5},m_{6}}
C^{j_{4}j_{5}j_{3}}_{m_{4},-n_{5},m_{3}}
C^{j_{4}j_{2}j_{6}}_{-n_{4},-n_{2},-n_{6}}\,,
\ee
where the $D^{j}(g)$ is the Wigner matrix representing the group element $g\in\SU(2)$ in the spin-$j$ representation.
This state is defined on the oriented dual graph, as illustrated on fig.\ref{spinnetwork}, with the dual vertex corresponding to the triangles and the dual link in one-to-one correspondance with the original edges.
This function is invariant under the action of $\SU(2)$ at each vertex of the dual graph:
\be
\vphi_{\{j_{i}\}}(\{g_{i}\})
=
\vphi_{\{j_{i}\}}(\{h_{s(i)}g_{i}h_{t(i)}^{-1}\})
\qquad\forall\quad h_{v}\in\SU(2)^{\times 4}
\,,
\ee
were $s(i)$ and $t(i)$ refer respectively to the source and target vertex of the dual link $i$ on the dual graph, and thus correspond to the two triangles sharing the edge $i$. The evaluation of this tetrahedral spin network wave-function at the identity $g_{i}=\id$ gives back the 6j-symbol:
\be
\vphi_{j_{1},..,j_{6}}(g_{1},..,g_{6})\bigg{|}_{g_{i}=\id}
\,=\,
\sixj{j_{1}}{j_{2}}{j_{3}}{j_{4}}{j_{5}}{j_{6}}\,.
\ee
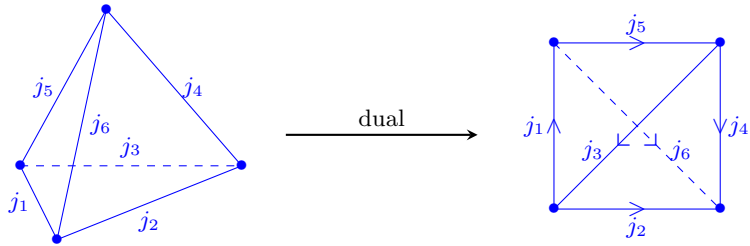
\begin{figure}[h!]
\centering
\begin{tikzpicture}[scale=1.7]
\coordinate (O1) at (0,0,0);

\coordinate (A1) at (0,1.061,0);
\coordinate (B1) at (0,-0.354,1);
\coordinate (C1) at (-0.866,-0.354,-0.5);
\coordinate (D1) at (0.866,-0.354,-0.5);

\draw[blue] (A1) -- (B1) node[midway,right]{$j_{6}$};
\draw[blue] (A1) -- (C1)node[midway,left]{$j_{5}$};
\draw[blue] (A1) -- (D1)node[midway,right]{$j_{4}$};
\draw[blue] (B1) -- (C1)node[midway,left]{$j_{1}$};
\draw[dashed,blue] (C1) -- (D1)node[midway,above]{$j_{3}$};
\draw[blue] (D1) -- (B1)node[midway,below]{$j_{2}$};

\draw[blue] (A1) node{$\bullet$};
\draw[blue] (B1) node{$\bullet$};
\draw[blue] (C1) node{$\bullet$};
\draw[blue] (D1) node{$\bullet$};

\draw[->,>=stealth, thick] (1.5,0.177,0.25) to node[midway,above]{dual} (3,0.177,0.25);

\coordinate (A2) at (3.5,0.8);
\coordinate (B2) at (3.5,-0.5);
\coordinate (C2) at (4.8,0.8);
\coordinate (D2) at (4.8,-0.5);
\draw[blue] (A2) node{$\bullet$};
\draw[blue] (B2) node{$\bullet$};
\draw[blue] (C2) node{$\bullet$};
\draw[blue] (D2) node{$\bullet$};

\draw[blue] (A2) -- (B2) node[midway]{$\w$} node[midway,left]{$j_{1}$} ;
\draw[blue] (A2) -- (C2) node[midway]{$>$} node[midway,above]{$j_{5}$};
\draw[dashed,blue] (A2) -- (D2) node[pos=0.58]{\large$\lrcorner$} node[pos=0.65,right]{$j_{6}$};
\draw[blue] (B2) -- (C2) node[pos=0.42]{\large$\llcorner$} node[pos=0.35,left]{$j_{3}$};
\draw[blue] (C2) -- (D2) node[midway]{$\vee$} node[midway,right]{$j_{4}$};
\draw[blue] (D2) -- (B2) node[midway]{$>$} node[midway,below]{$j_{2}$};

%
%

\end{tikzpicture}

\caption{From the tetrahedron to the dual spin network graph: each triangle represented by a dual vertex and each edge by a dual link, the  dual graph also has the combinatorics of a tetrahedron.}
\label{spinnetwork}
\end{figure}
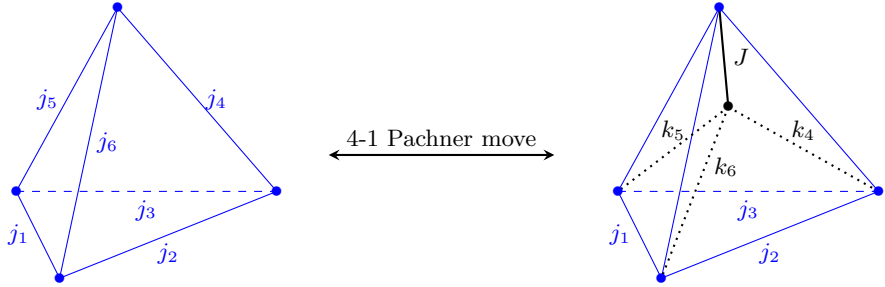
\begin{figure}[h!]
\centering
\begin{tikzpicture}[scale=2]

\coordinate (A1) at (0,1.061,0);
\coordinate (B1) at (0,-0.354,1);
\coordinate (C1) at (-0.866,-0.354,-0.5);
\coordinate (D1) at (0.866,-0.354,-0.5);

\draw[blue] (A1) -- (B1) node[midway,right]{$j_{6}$};
\draw[blue] (A1) -- (C1)node[midway,left]{$j_{5}$};
\draw[blue] (A1) -- (D1)node[midway,right]{$j_{4}$};
\draw[blue] (B1) -- (C1)node[midway,left]{$j_{1}$};
\draw[dashed,blue] (C1) -- (D1)node[midway,below]{$j_{3}$};
\draw[blue] (D1) -- (B1)node[midway,below]{$j_{2}$};

\draw[blue] (A1) node{$\bullet$};
\draw[blue] (B1) node{$\bullet$};
\draw[blue] (C1) node{$\bullet$};
\draw[blue] (D1) node{$\bullet$};

\coordinate (O2) at (4.06,0.4,0);

\coordinate (A2) at (4,1.061,0);
\coordinate (B2) at (4,-0.354,1);
\coordinate (C2) at (3.134,-0.354,-0.5);
\coordinate (D2) at (4.866,-0.354,-0.5);

\draw[blue] (A2) -- (B2) ;
\draw[blue] (A2) -- (C2);
\draw[blue] (A2) -- (D2);
\draw[blue] (B2) -- (C2)node[midway,left]{$j_{1}$};
\draw[dashed,blue] (C2) -- (D2)node[midway,below]{$j_{3}$};
\draw[blue] (D2) -- (B2)node[midway,below]{$j_{2}$};

\draw[thick] (O2) -- (A2)node[midway,right]{$J$};
\draw[dotted, thick] (O2) -- (B2)node[pos=0.35,right]{$k_{6}$};
\draw[dotted, thick] (O2) -- (C2)node[midway,above]{$k_{5}$};
\draw[dotted, thick] (O2) -- (D2)node[midway,above]{$k_{4}$};

\draw (O2) node{$\bullet$};
\draw[blue] (A2) node{$\bullet$};
\draw[blue] (B2) node{$\bullet$};
\draw[blue] (C2) node{$\bullet$};
\draw[blue] (D2) node{$\bullet$};

\draw[<->,>=stealth, thick] (1.5,0.177,0.25) to node[midway,above]{4-1 Pachner move} (3,0.177,0.25);

\end{tikzpicture}

\caption{The holonomy operator around the dual cycle $(456)$ corresponding to a summit of the tetrahedron induces a tent move moving the summit and thus shifting the spins $j_{4},j_{5},j_{6}$, which can be interpreted as a 4-1 Pachner move creating a vertex within the tetrahedron.}
\label{tent}
\end{figure}
Following \cite{Bonzom:2009zd}, we derive the finite 4-1 Pachner move identity from the action of holonomy operator on a 3-cycle of the dual graph. Let's act on the cycle $(4,5,6)$ corresponding to a summit of the tetrahedron. This leads to a ``tent''-move leaving the spins $j_{1},j_{2},j_{3}$ on the ground unchanged and shifting the spins $j_{4},j_{5},j_{6}$, according to fig.\ref{tent}. Mathematically, we use spin recoupling formulas to turn tensor products into Clebsch-Gordan coefficients and then into 6j-symbols to finally get:
\begin{multline} 
\chi_{J}(g_{4}g_{6}^{-1}g_{5})\vphi_{j_{1},..,j_{6}}(\{g_{i}\})
=
\sum_{k_{4},k_{5},k_{6}}
(-1)^{\sum_{i}j_{i}+k_{4}+k_{5}+k_{6}+J}d_{k_{4}}d_{k_{5}}d_{k_{6}}\\
\times\sixj{k_{4}}{j_{4}}{J}{j_{6}}{k_{6}}{j_{2}}
\sixj{k_{5}}{j_{5}}{J}{j_{4}}{k_{4}}{j_{3}}
\sixj{k_{6}}{j_{6}}{J}{j_{5}}{k_{5}}{j_{1}}
\vphi_{j_{1},j_{2},j_{3},k_{4},k_{5},k_{6}}(\{g_{i}\})\,.
\end{multline} 
We now evaluate this state at the identity $g_{i}=\id$, which can be understood as the flat connection state on the tetrahedron:
\be
\langle\id|\widehat{\chi_{J}^{(456)}}|\vphi_{\{j_{i}\}}\rangle
=
\chi_{J}(\id)\vphi_{j_{1},..,j_{6}}(\{g_{i}\})\bigg{|}_{g_{i}=\id}
=d_{J}\,\sixj{j_{1}}{j_{2}}{j_{3}}{j_{4}}{j_{5}}{j_{6}}\,,
\ee
where $d_{J}=\chi_{J}(\id)=(2J+1)$ is the dimension of the spin-$J$ representation.
As explained in \cite{Bonzom:2009zd}, this provides a finite identity for the 4-1 Pachner move\footnotemark{}:
\begin{multline} 
d_{J}\,\sixj{j_{1}}{j_{2}}{j_{3}}{j_{4}}{j_{5}}{j_{6}}=
\sum_{k_{4},k_{5},k_{6}}
(-1)^{\sum_{i}j_{i}+k_{4}+k_{5}+k_{6}+J}d_{k_{4}}d_{k_{5}}d_{k_{6}}\\
\times\sixj{k_{4}}{j_{4}}{J}{j_{6}}{k_{6}}{j_{2}}
\sixj{k_{5}}{j_{5}}{J}{j_{4}}{k_{4}}{j_{3}}
\sixj{k_{6}}{j_{6}}{J}{j_{5}}{k_{5}}{j_{1}}
\sixj{j_{1}}{j_{2}}{j_{3}}{k_{4}}{k_{5}}{k_{6}}\,.
\label{41move}
\end{multline}
This means that the Ponzano-Regge amplitude for the finer triangulation $\Delta$ with an extra vertex and fixing one of the spins, here $J$, on an edge attached to that vertex is equal to the amplitude on the coarser triangulation $\Delta_{0}$ without that vertex, up to a simple factor depending on $J$:
\be
\cZ^{PR}_{\Delta}(J)
=
d_{J}^{2}\,\cZ^{PR}_{\Delta_{0}}\,.
\ee
This realizes the invariance of the Ponzano-Regge model under the 4-1 Pachner move.
This procedure allows to relate explicitly the Hamiltonian constraint operators for 3d quantum gravity, acting as holonomy operators, to the topological invariance of the Ponzano-Regge model and its coarse-graining.
\footnotetext{One can show that this identity actually follows from the Biedenharn-Elliot identity for the 3-2 Pachner move. Indeed, we combine \eqref{32move}  with the orthonormality relation satisfied by the 6j-symbol:
$$
\sum_{k} d_{k}
\sixj{k_{6}}{j_{2}}{k}{j_{3}}{k_{5}}{p}
\sixj{j_{3}}{k_{5}}{k}{k_{6}}{j_{2}}{q}
=
\f{\delta_{pq}}{d_{p}}
\,.
$$
This leads to a stronger identity involving the sums over two spins, say $k_{4}$ and $k_{5}$. For all spins $k_{6}$ between $|j_{6}-J|$ and $j_{6}+J$, we have the equality:
$$
\sum_{k_{4},k_{5}}
(-1)^{\sum_{i}j_{i}+k_{4}+k_{5}+k_{6}+J}d_{k_{4}}d_{k_{5}}\\
\times\sixj{k_{4}}{j_{4}}{J}{j_{6}}{k_{6}}{j_{2}}
\sixj{k_{5}}{j_{5}}{J}{j_{4}}{k_{4}}{j_{3}}
\sixj{k_{6}}{j_{6}}{J}{j_{5}}{k_{5}}{j_{1}}
\sixj{j_{1}}{j_{2}}{j_{3}}{k_{4}}{k_{5}}{k_{6}}
\,=\,
\f1{d_{j_{6}}}\,\sixj{j_{1}}{j_{2}}{j_{3}}{j_{4}}{j_{5}}{j_{6}}
\,.
$$
From this identity we can deduce the 4-1 Pachner move \eqref{41move} by summing over $k_{6}$ and using the well-known dimension sum formula for spin tensor products:
$$
\sum_{k_{6}=|j_{6}-J|}^{j_{6}+J}d_{k_{6}}\,=\,d_{j_{6}}d_{J}\,.
$$
}
A last remark on the 4-1 Pachner move is than the un-gauge-fixed Ponzano-Regge amplitude $\cZ^{PR}_{\Delta}$ on the finer triangulation $\Delta$ is simply the sum over the bulk spin $J$ of our gauge-fixed amplitude $\cZ^{PR}_{\Delta}(J)$. This leads to the expected divergence of the Ponzano-Regge model due to a bulk vertex (in the triangulation):
\be
\cZ^{PR}_{\Delta}=\sum_{J}\cZ^{PR}_{\Delta}(J)=\Big{(}\sum_{J}d_{J}^{2}\Big{)}\,\cZ^{PR}_{\Delta_{0}}
=\delta(\id)\,\cZ^{PR}_{\Delta_{0}}\,.
\ee
Fixing a bulk spin in order to gauge-fix, and thus regularize, the Ponzano-Regge amplitude is exactly the procedure considered in \cite{Freidel:2004vi}. However this previous work focused on the case $J=0$, when the gauge-fixing is equivalent to collapsing the triangulation and effectively removing the bulk vertices, while we consider here the more general case of gauge-fixing a bulk spin to an arbitrary value $J$.

\subsection{Higher derivatives and coarse-graining of edge lengths}

We have seen how the Ponzano-Regge partition function is explicitly invariant under coarse-graining, through a finite identity satisfied by the 6j-symbol implementing the 4-1 Pachner move. Next, we are interested in the behavior of geometric observables, such as areas and volumes. In this section, we derive new identities on the 6j-symbol describing the propagation of area and volume observables by a 4-1 Pachner move.

The method consists in two simple steps. First we use that the 4-1 Pachner move identity is realized by applying the holonomy operator to a tetrahedral spin network state and then evaluating the resulting state against the flat state. Second, we follow the suggestion made in \cite{Charles:2016xwc} to investigate higher derivatives of the flat state, defined as (gauge-invariant- grasping operators applied to the $\delta$-distribution peaked on flat holonomies. Overall, this amounts to inserting grasping operators in the scalar product $\langle\id|\widehat{\chi_{J}^{(456)}}|\vphi_{\{j_{i}\}}\rangle$.

Let us start with a $\vJ^{(2)}\cdot \vJ^{(4)}$ insertion\footnotemark. On the one hand, we can evaluate that scalar product by an integration by parts:
\footnotetext{Since $\SU(2)$ is non-abelian, we have left and right derivative $\pp_{g}^{L}$ and $\pp_{g}^{R}$ for each variable $g\in\SU(2)$. Here we consider gauge-invariant differential operators. Quadratic grasping operators are attached to vertices of the dual graph, so that left/right subscript are obvious and can be kept implicit\pounds. For instance $\vJ^{(2)}\cdot \vJ^{(4)}$ stands for $\vJ^{(2)R}\cdot \vJ^{(4)R}$, and so on for $\vJ^{(3)L}\cdot \vJ^{(4)L}$, $\vJ^{(1)R}\cdot \vJ^{(5)L}$ and $\vJ^{(4)L}\cdot \vJ^{(5)R}$. }
\beq
\langle\id|\vJ^{(2)}\cdot \vJ^{(4)}\,\widehat{\chi_{J}^{(456)}}|\vphi_{\{j_{i}\}}\rangle
&=&
\int [\rd g_{i}]^{\times 6}\,
\left[(\vJ^{(2)}\cdot \vJ^{(4)})\,\delta(g_{1}g_{6}g_{2}^{-1})\delta(g_{1}g_{5}g_{3})\delta(g_{2}g_{4}^{-1}g_{3})\right]\,
\chi_{J}(g_{4}g_{6}^{-1}g_{5})\vphi_{\{j_{i}\}}(g_{1},..,g_{6})\nn\\
&=&
\int [\rd g_{i}]^{\times 6}\,
\delta(g_{1}g_{6}g_{2}^{-1})\delta(g_{1}g_{5}g_{3})\delta(g_{2}g_{4}^{-1}g_{3})\,
\chi_{J}(g_{4}g_{6}^{-1}g_{5})\,
\left[(\vJ^{(2)}\cdot \vJ^{(4)})\,\vphi_{\{j_{i}\}}(g_{1},..,g_{6})\right]\nn\\
&=&
\f{d_{J}}2\Big{[}
j_{6}(j_{6}+1)-j_{2}(j_{2}+1)-j_{4}(j_{4}+1)
\Big{]}
\,\sixj{j_{1}}{j_{2}}{j_{3}}{j_{4}}{j_{5}}{j_{6}}\,.
\eeq
The  three $\delta$-distributions constrain the holonomies around independent loops on the dual graph to be trivial. This is enough to ensure that gauge-invariant functions are evaluated at the identity $g_{i}=\id$ for all $i=1..6$.
On the other hand, one can apply directly the holonomy on the spin network state $\vphi_{\{j_{i}\}}$, proceeding to the 4-1 Pachner move, and then apply the grasping and evaluate the resulting state at the identity:
\begin{multline} 
\langle\id|\vJ^{(2)}\cdot \vJ^{(4)}\,\widehat{\chi_{J}^{(456)}}|\vphi_{\{j_{i}\}}\rangle=
\sum_{k_{4},k_{5},k_{6}}
(-1)^{\sum_{i}j_{i}+k_{4}+k_{5}+k_{6}+J}d_{k_{4}}d_{k_{5}}d_{k_{6}}\,
\f12\Big{[}
k_{6}(k_{6}+1)-j_{2}(j_{2}+1)-k_{4}(k_{4}+1)\Big{]}
\\
\times\sixj{k_{4}}{j_{4}}{J}{j_{6}}{k_{6}}{j_{2}}
\sixj{k_{5}}{j_{5}}{J}{j_{4}}{k_{4}}{j_{3}}
\sixj{k_{6}}{j_{6}}{J}{j_{5}}{k_{5}}{j_{1}}
\sixj{j_{1}}{j_{2}}{j_{3}}{k_{4}}{k_{5}}{k_{6}}\,.
\end{multline}
Ignoring the Casimir term $j_{2}(j_{2}+1)$ which is actually not involved in the tent move, we have shown that the geometric observable $k_{6}(k_{6}+1)-k_{4}(k_{4}+1)$ propagates trivially under coarse-graining by the 4-1 Pachner move:
\begin{multline} 
{d_{J}}\Big{[}
j_{6}(j_{6}+1)-j_{4}(j_{4}+1)
\Big{]}
\,\sixj{j_{1}}{j_{2}}{j_{3}}{j_{4}}{j_{5}}{j_{6}}=
\sum_{k_{4},k_{5},k_{6}}
(-1)^{\sum_{i}j_{i}+k_{4}+k_{5}+k_{6}+J}d_{k_{4}}d_{k_{5}}d_{k_{6}}\,
\Big{[}
k_{6}(k_{6}+1)-k_{4}(k_{4}+1)\Big{]}
\\
\times\sixj{k_{4}}{j_{4}}{J}{j_{6}}{k_{6}}{j_{2}}
\sixj{k_{5}}{j_{5}}{J}{j_{4}}{k_{4}}{j_{3}}
\sixj{k_{6}}{j_{6}}{J}{j_{5}}{k_{5}}{j_{1}}
\sixj{j_{1}}{j_{2}}{j_{3}}{k_{4}}{k_{5}}{k_{6}}\,.
\end{multline}
One can proceed similarly with other quadratic grasping operators, such as $(\vJ^{(4)})^{2}$ or $\vJ^{(4)}\cdot\vJ^{(6)}$.  Nevertheless, for these insertion, the integration by parts will produce an extra-term:
$$
\big{(}\vJ^{(4)}\big{)}^{2}\,\chi_{J}(g_{4}g_{6}^{-1}g_{5})\bigg{|}_{g_{i}=\id}
\,=\,
J(J+1)\,d_{J}\,,
$$
and similarly for $\vJ^{(4)}\cdot\vJ^{(6)}$. This leads to other identities, giving the explicit transformation of the squared length observables  $k_{4}(k_{4}+1)$, $k_{5}(k_{5}+1)$ and $k_{6}(k_{6}+1)$:
\begin{multline} 
{d_{J}}\Big{[}
j_{4}(j_{4}+1)+J(J+1)
\Big{]}
\,\sixj{j_{1}}{j_{2}}{j_{3}}{j_{4}}{j_{5}}{j_{6}}=
\sum_{k_{4},k_{5},k_{6}}
(-1)^{\sum_{i}j_{i}+k_{4}+k_{5}+k_{6}+J}d_{k_{4}}d_{k_{5}}d_{k_{6}}\,
\Big{[}
k_{4}(k_{4}+1)\Big{]}
\\
\times\sixj{k_{4}}{j_{4}}{J}{j_{6}}{k_{6}}{j_{2}}
\sixj{k_{5}}{j_{5}}{J}{j_{4}}{k_{4}}{j_{3}}
\sixj{k_{6}}{j_{6}}{J}{j_{5}}{k_{5}}{j_{1}}
\sixj{j_{1}}{j_{2}}{j_{3}}{k_{4}}{k_{5}}{k_{6}}\,.
\label{pachner41length}
\end{multline}
In operator terms, this formula computes the commutator of the holonomy operator, interpreted as a Hamiltonian constraint, and the Casimir operator $(\vJ^{(4)}\big{)}^{2}$ when evaluated on a physical state $|\id\rangle$. This commutator does not vanish, because the Casimir operator $(\vJ^{(4)}\big{)}^{2}$ is not a physical observable - in the sense that it does not commute with the Hamiltonian constraints (quantized as the holonomy operators), or equivalently it is not invariant under the translational symmetry of 3d gravity. But the identity on the 6j-symbol allows to keep this commutator under control:
\be
\langle\id|\,\bigg{[}(\vJ^{(4)}\big{)}^{2}\,,\,\widehat{\chi_{J}^{(456)}}\,\bigg{]}
\,|\vphi_{\{j_{i}\}}\rangle
\,=\,
\cC_{J}\,
\langle\id|\,\widehat{\chi_{J}^{(456)}}
\,|\vphi_{\{j_{i}\}}\rangle
\,,
\ee
where we recall that $\cC_{J}=J(J+1)$ is the value of the $\SU(2)$ Casimir for the spin-$J$ representation.

\medskip

Although this formula is non-trivial algebraic result, it is fairly direct to endow it with a semi-classical interpretation in terms of geometry. Considering the picture \eqref{tent} for the tent move induced by the holonomy operator $\widehat{\chi_{J}^{(456)}}$, the Casimir operator $k_{4}(k_{4}+1)$ corresponds to the squared length of a bulk edge, defined by say the vector $\vw_{4}$, while the Casimir operator $j_{4}(j_{4}+1)$ corresponds to the squared length of a boundary edge, say $\vv_{4}$. The 4-1 Pachner move consists in displacing the summit of the tetrahedron by a vector $J\hat{u}$ of norm given by $J$ with arbitrary unit direction $\hat{u}$ living on the 2-sphere. The sum over the bulk spins $k_{4},k_{5},k_{6}$ amounts to the integration over all possible displacements $\hat{u}\in\cS^{2}$, thus leading to:
\be
\int_{\cS^{2}}\rd^{2}\hat{u}\,
|\vw_{4}|^{2}
\,=\,
\int_{\cS^{2}}\rd^{2}\hat{u}\,
|\vv_{4}+J\hat{u}|^{2}
\,=\,
|\vv_{4}|^{2}+J^{2}\,,
\ee
which is the classical geometry counterpart of the identity \eqref{pachner41length} above on the 6j-symbols. This might seem counter-intuitive at first, since $|\vw_{4}|$ seems shorter than $|\vv_{4}|$ in fig.\ref{tent}, but the new bulk vertex can actually be inside or outside the original tetrahedron.

\subsection{Triple grasping and coarse-graining of the tetrahedron volume}

We have investigated up the coarse-grained of the individual spins, reflecting the edge lengths. The next basic geometrical observable is the tetrahedron volume. At the quantum level, it is implemented by triple grasping operators as differential operators acting on three edges of the tetrahedron. Let us consider the triple grasping (Hermitian) operator $\hat{T}_{456}\,\equiv\,i\vJ^{(4)}\cdot(\vJ^{(5)}\w\vJ^{(6)})$. As explained in \cite{Hackett:2006gp} and as checked in appendix \ref{V3}, this operator does not give straightforwardly the volume in the large spin semi-classical limit due to a simple $\cos$ versus $\sin$ substitution but its square $\hat{V}^{2}$ does allow to recover the classical squared volume of the tetrahedron.

The explicit action of the triple grasping on the 6j-symbol is computed in the appendix of \cite{Hackett:2006gp}, which we give here for the sake of completeness:
\beq
\hat{T}_{456}\sixj{j_{1}}{j_{2}}{j_{3}}{j_{4}}{j_{5}}{j_{6}}
&=&
-\sqrt{6\prod_{i=4,5,6}d_{j_{i}}\cC_{j_{i}}}\,
\sum_{k}
d_{k}\sixj{j_{3}}{j_{4}}{j_{5}}{1}{j_{5}}{k}
\sixj{j_{2}}{j_{4}}{j_{6}}{1}{j_{6}}{k}
\sixj{j_{4}}{j_{4}}{1}{1}{1}{k}
\sixj{j_{1}}{j_{2}}{j_{3}}{k}{j_{5}}{j_{6}} 
\nn\\
&=&
\alpha_{-}\sixj{j_{1}}{j_{2}}{j_{3}}{j_{4}-1}{j_{5}}{j_{6}}
+\alpha_{0}\sixj{j_{1}}{j_{2}}{j_{3}}{j_{4}}{j_{5}}{j_{6}}
+\alpha_{+}\sixj{j_{1}}{j_{2}}{j_{3}}{j_{4}+1}{j_{5}}{j_{6}}\,,
\label{T456}\eeq
with $\cC_{j}=j(j+1)$ the value of the $\SU(2)$ Casimir for a spin $j$ and the $\alpha$-coefficients given as:
\begin{flalign}
\alpha_{-}
\,=\,
\f{(j_{4}+1)}{4j_{4}(2j_{4}+1)}
&\sqrt{(j_{4}+j_{5}-j_{3})(j_{4}-j_{5}+j_{3})(1-j_{4}+j_{5}+j_{3})(1+j_{4}+j_{5}+j_{3})}&
\nn\\
&\times
\sqrt{(j_{4}+j_{2}-j_{6})(j_{4}-j_{2}+j_{6})(1-j_{4}+j_{2}+j_{6})(1+j_{4}+j_{2}+j_{6})}\,,&
\end{flalign}
\begin{flalign}
\alpha_{0}
\,=\,
-\f{\Big{[}\cC_{j_{4}}+\cC_{j_{5}}-\cC_{j_{3}}\Big{]}\Big{[}\cC_{j_{4}}+\cC_{j_{2}}-\cC_{j_{6}}\Big{]}}{4\cC_{j_{4}}}\,,
&&
\end{flalign}
\begin{flalign}
\alpha_{+}
\,=\,
\f{j_{4}}{4(j_{4}+1)(2j_{4}+1)}
&\sqrt{(1+j_{4}+j_{5}-j_{3})(1+j_{4}-j_{5}+j_{3})(-j_{4}+j_{5}+j_{3})(2+j_{4}+j_{5}+j_{3})}&
\nn\\
&\times
\sqrt{(1+j_{4}+j_{2}-j_{6})(1+j_{4}-j_{2}+j_{6})(-j_{4}+j_{2}+j_{6})(2+j_{4}+j_{2}+j_{6})}\,.&
\end{flalign}

\medskip

Then, on the one hand, it is rather direct to compute the scalar product $\langle\id|\hat{T}_{456}\,\widehat{\chi_{J}^{(456)}}|\vphi_{\{j_{i}\}}\rangle$ by integration by parts:
\be
\langle\id|\hat{T}_{456}\,\widehat{\chi_{J}^{(456)}}|\vphi_{\{j_{i}\}}\rangle
\,=\,
d_{J}\,\Big{(}
\langle\id|\hat{T}_{456}\,|\vphi_{\{j_{i}\}}\rangle-J(J+1) \langle\id|\vphi_{\{j_{i}\}}\rangle\Big{)}
\,=\,
d_{J}\,\hat{T}_{456}\sixj{j_{1}}{j_{2}}{j_{3}}{j_{4}}{j_{5}}{j_{6}}
-d_{J}J(J+1)\sixj{j_{1}}{j_{2}}{j_{3}}{j_{4}}{j_{5}}{j_{6}}\,,
\ee
where $\hat{T}_{456}\{6j\}$ stands for the triple grasping acting on the 6j-symbol. Indeed, only two terms survive the integration by parts: either all the derivatives hit the spin network function $\vphi_{\{j_{i}\}}$ or they all hit the holonomy insertion $\chi_{J}(g_{4}g_{6}^{-1}g_{5})$. The latter case produce a Casimir factor:
$$
i\vJ^{(4)}\cdot(\vJ^{(5)}\w\vJ^{(6)})\chi_{J}(g_{4}g_{6}^{-1}g_{5})\bigg{|}_{g_{i}=\id}
\,=\,
i\chi_{J}(\eps_{ijk}J_{i}J_{j}J_{k})
\,=\,
-\chi_{J}(J_{i}J_{i})
\,=\,
-J(J+1)\,\chi_{J}(\id)
\,=\,
-J(J+1)\,d_{J}\,.
$$
This implies the following identity about the 4-1 Pachner move transformation of the triple grasping:
\begin{multline} 
{d_{J}}\hat{T}_{456}\sixj{j_{1}}{j_{2}}{j_{3}}{j_{4}}{j_{5}}{j_{6}}
-{d_{J}}J(J+1)\,\sixj{j_{1}}{j_{2}}{j_{3}}{j_{4}}{j_{5}}{j_{6}}
=
\sum_{k_{4},k_{5},k_{6}}
(-1)^{\sum_{i}j_{i}+k_{4}+k_{5}+k_{6}+J}d_{k_{4}}d_{k_{5}}d_{k_{6}}\,
\\
\times\sixj{k_{4}}{j_{4}}{J}{j_{6}}{k_{6}}{j_{2}}
\sixj{k_{5}}{j_{5}}{J}{j_{4}}{k_{4}}{j_{3}}
\sixj{k_{6}}{j_{6}}{J}{j_{5}}{k_{5}}{j_{1}}\,
\hat{T}_{456}\sixj{j_{1}}{j_{2}}{j_{3}}{k_{4}}{k_{5}}{k_{6}}\,.
\label{pachner41T456}
\end{multline}
which is effectively more algebraically-involved but keeps a simple structure. This new identity has been thoroughly checked numerically using the explicit formula \eqref{T456} for the triple grasping through Mathematica simulations.

\medskip

The non-trivial triple grasping operators acting on the 6j-symbol are all equivalent to one of the four triple grasping acting on the edges around a tetrahedron summit, or equivalently a 3-cycle of the dual graph, that $\hat{T}_{456}$, $\hat{T}_{234}$, $\hat{T}_{135}$ and $\hat{T}_{126}$. Besides the triple grasping  $\hat{T}_{456}$ considered above, the other three triple graspings propagate trivially under the 4-1 Pachner move and do not get any correction term. We can then introduce a symmetrized volume operator as:
\be
\hat{V}=\f14
\left(\hat{T}_{456}+\hat{T}_{234}+\hat{T}_{135}+\hat{T}_{126}
\right)\,,
\ee
which leads to the final 4-1 Pachner move identity:
\begin{multline} 
{d_{J}}\hat{V}\sixj{j_{1}}{j_{2}}{j_{3}}{j_{4}}{j_{5}}{j_{6}}
-{d_{J}}\f{J(J+1)}4\,\sixj{j_{1}}{j_{2}}{j_{3}}{j_{4}}{j_{5}}{j_{6}}
=
\sum_{k_{4},k_{5},k_{6}}
(-1)^{\sum_{i}j_{i}+k_{4}+k_{5}+k_{6}+J}d_{k_{4}}d_{k_{5}}d_{k_{6}}\,
\\
\times\sixj{k_{4}}{j_{4}}{J}{j_{6}}{k_{6}}{j_{2}}
\sixj{k_{5}}{j_{5}}{J}{j_{4}}{k_{4}}{j_{3}}
\sixj{k_{6}}{j_{6}}{J}{j_{5}}{k_{5}}{j_{1}}\,
\hat{V}\sixj{j_{1}}{j_{2}}{j_{3}}{k_{4}}{k_{5}}{k_{6}}\,.
\label{pachner41volume}
\end{multline}
In operator terms, this formula computes the commutator of the holonomy operator, interpreted as a Hamiltonian constraint, and the triple graspings when evaluated on a physical state $|\id\rangle$:
\be
\langle\id|\,\bigg{[}\hat{T}_{456}\,,\,\widehat{\chi_{J}^{(456)}}\,\bigg{]}
\,|\vphi_{\{j_{i}\}}\rangle
\,=\,
-\cC_{J}\,
\langle\id|\,\widehat{\chi_{J}^{(456)}}
\,|\vphi_{\{j_{i}\}}\rangle
\,,
\qquad
\langle\id|\,\bigg{[}\hat{V}\,,\,\widehat{\chi_{J}^{(456)}}\,\bigg{]}
\,|\vphi_{\{j_{i}\}}\rangle
\,=\,
-\f14\cC_{J}\,
\langle\id|\,\widehat{\chi_{J}^{(456)}}
\,|\vphi_{\{j_{i}\}}\rangle
\,.
\ee
This concludes our analysis of the behavior of the length and volume observables under coarse-graining by 4-1 Pachner move in the Ponzano-Regge model and the associated new identities satisfied by the 6j-symbol.  We have seen that these geometric observables somehow almost propagate trivially under the 4-1 Pachner move up to J-dependent correction terms which can be computed explicitly (and derived simply by an integration by parts).

\section{Expanding the $q$-deformed 6j-symbol}

The new formulas for the 4-1 Pachner move, which we derived in the previous section, satisfied by the 6j-symbol with length and volume insertions hint towards the possibility of identifying further solutions to the invariance under 4-1 Pachner moves besides the 6j-symbol itself. This would provide new models, either directly invariant under coarse-graining or at least with a well-defined and controlled behavior under coarse-graining. This would be a very interesting arena for the study of the coarse-graining of spinfoam models and investigating the interplay between (Hamiltonian) dynamics and coarse-graining for quantum gravity.

Actually, we already know of a whole class of other symbols invariant under Pachner moves and leading to topological models: these are the $q$-deformed 6j-symbols $\{6j\}_{q}$, leading to the Turaev-Viro topological invariant for $q$ root of unity \cite{Turaev:1992hq} or its hyperbolic counterpart for $q$ real \cite{Dupuis:2013haa,Bonzom:2014bua}. So we show below that the new identities that we derived on the behavior of the lengths and volume under coarse-graining imply the invariance of the $\{6j\}_{q}$-symbols under 4-1 Pachner moves at leading order in $q$. In this light, we postpone the development of new 3d models which controlled (and possibly non-trivial) behavior under 4-1 Pachner moves to future investigation.

\subsection{The volume formula for the $q$-deformation at leading order}

\begin{figure}[h!]


\begin{subfigure}[t]{.45\linewidth}
\includegraphics[scale=0.5]{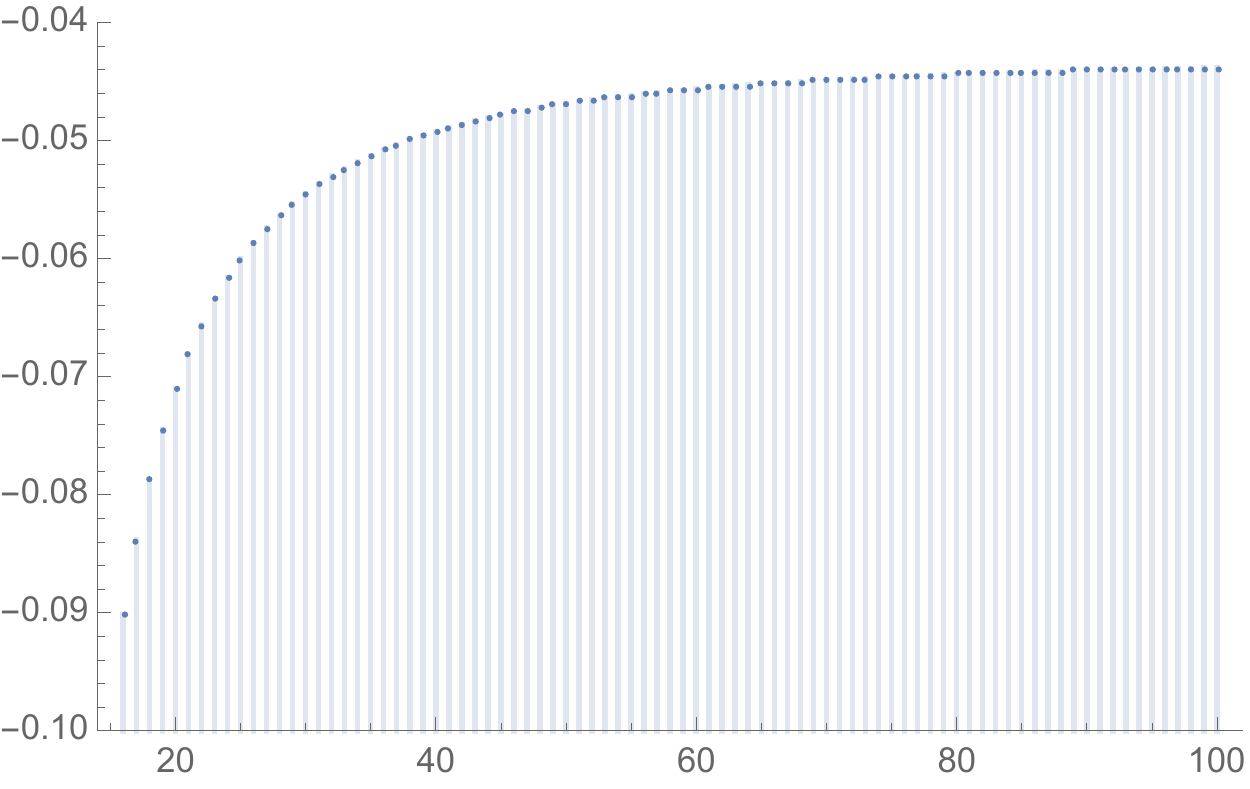}
\caption{The $\{6j\}_{q}$-symbol for $j_{i}=2$ and $q=\exp \f {2i\pi}k$ as $k$ varies from 10 to 100.}
\end{subfigure}
\hspace{2mm}
\begin{subfigure}[t]{.45\linewidth}
\includegraphics[scale=0.5]{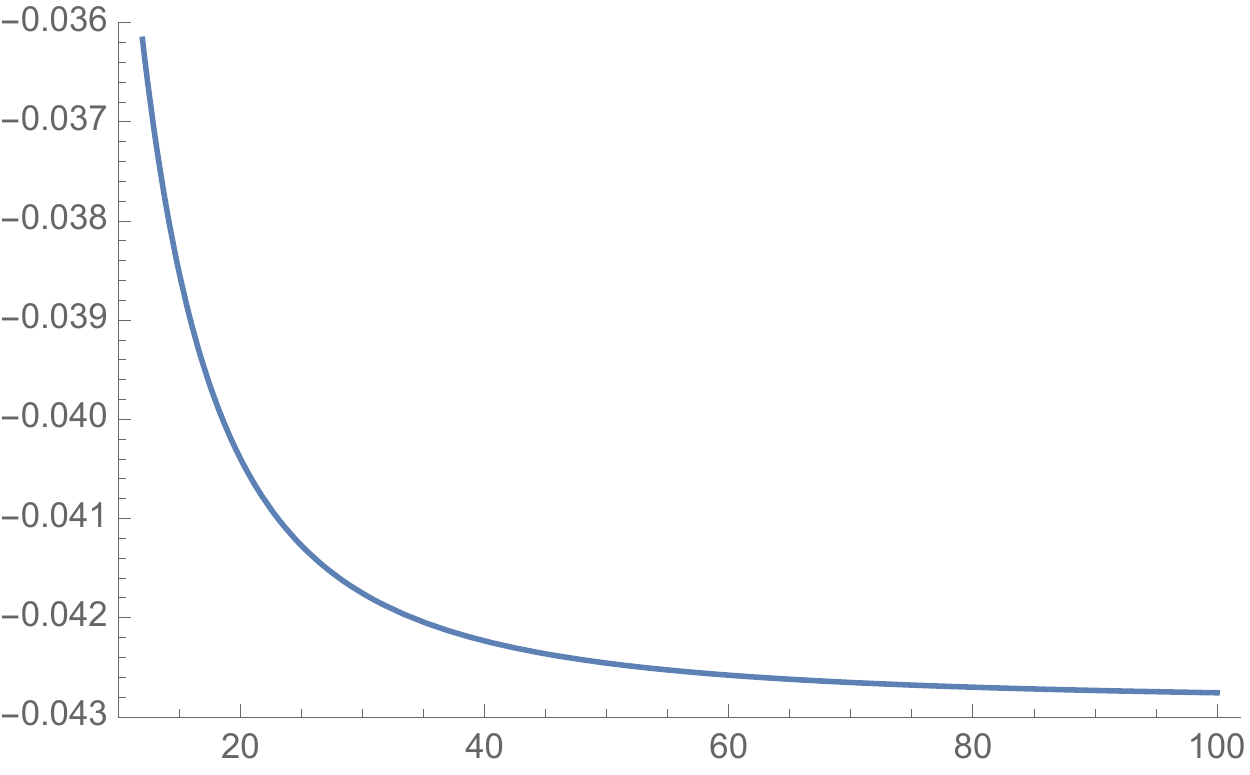}
\caption{The $\{6j\}_{q}$-symbol for $j_{i}=2$ and $q=\exp 2\tau$ as $\tau^{-1}$ varies from 10 to 100.}
\end{subfigure} 

\caption{The $\{6j\}_{q}$-symbol for $j_{i}=2$ in terms of the deformation parameter $q$: as $q$ goes to 1, we recover the classical limit $-\f3{70}$ given by the standard 6j-symbol.}
\label{fig:6jq}

\end{figure}

The $q$-deformed 6j-symbol is defined in terms of $q$-numbers:
\be
[n]_{q}
=
\f{q^{\f n2}-q^{-\f n2}}{q^{\f 12}-q^{-\f 12}}
\,,\quad
\forall n\in\N\,.
\ee
In the classical limit $q\rightarrow 1$, the $q$-integer $[n]_{q}$ go back to $n$. But in the general case, these $q$-integers have a different behavior if $q$ is chosen as a root of unity $q\in\U(1)$ or as a real number $q\in\R_{+}$:
\be
q=e^{i\f{2\pi}k}
\quad
\Rightarrow
\quad
[n]_{q}
=
\f{\sin \f{\pi n}k}{\sin \f{\pi}k}\,,
\qquad\qquad
q=e^{2\tau}
\quad
\Rightarrow
\quad
[n]_{q}
=
\f{\sinh n{ \tau}}{\sinh\tau}\,.
\ee
In the context of 3d quantum gravity (with Euclidean signature), the unitarity case corresponds to a positive cosmological constant $\Lambda\ge0$, with $q$-deformation $q=\exp i\sqrt{\Lambda}$ while the real case corresponds to a negative cosmological constant $\Lambda\le0$ with deformation parameter $q=\exp \sqrt{-\Lambda}$. In both cases, the $q$-integers have the same infinitesimal behavior close to $\Lambda\rightarrow 0$, $q\rightarrow 1$:
\be
\pp_{\Lambda}[n]_{q}\bigg{|}_{\Lambda\rightarrow0}\,=\,\f{-1}{24}(n-1)n(n+1)\,.
\ee
The $q$-dimension of the representation of spin-$j$ is given by the $q$-version of the standard dimension:
\be
\dim_{q} j = [2j+1]_{q}\,,
\qquad
\pp_{\Lambda}[2j+1]_{q}\bigg{|}_{\Lambda\rightarrow0}\,=\,\f{-1}{6}\,(2j+1)\,j(j+1)
\,,
\label{qdim}
\ee
where the $\SU(2)$ Casimir appears. We can further introduce $q$-factorials and the $q$-version of the triangular coefficients:
\be
[n]_{q}!=\prod_{m=1}^{n}[m]_{q}\,,
\qquad
\Delta_{abc}^{(q)}
=
\f{[-a+b+c]_{q}![a-b+c]_{q}![a+b-c]_{q}!}{[a+b+c+1]_{q}!}\,,
\ee
where $a,b,c$ satisfy the triangular inequalities and $(a+b+c+1)$ is taken modulo $k$ in the root of unity case.
Finally the $\{6j\}_{q}$-symbol can be computed as a deformed Racah sum as in the standard case:
\be
\sixj{j_{1}}{j_{2}}{j_{3}}{j_{4}}{j_{5}}{j_{6}}_{q}
\,=\,
\sqrt{\Delta_{j_{1}j_{2}j_{3}}^{(q)}\Delta_{j_{1}j_{5}j_{6}}^{(q)}\Delta_{j_{4}j_{2}j_{6}}^{(q)}\Delta_{j_{4}j_{5}j_{3}}^{(q)}}\,
\sum_{t=\max{\alpha_{k}}}^{\min{\beta_{l}}}
(-1)^{t}
\f{[t+1]_{q}!}{\prod_{k=1}^{4}[t-\alpha_{k}]_{q}!\,\prod_{l=1}^{3}[\beta_{l}-t]_{q}!}
\ee
where the $\alpha$'s and $\beta$'s are respectively the sum of spins around the 3- and 4-cycles of the tetrahedron:
$$
\alpha_{k}=
\left|
\begin{array}{l}
j_{1}+j_{2}+j_{3}\\
j_{1}+j_{5}+j_{6}\\
j_{4}+j_{2}+j_{6}\\
j_{4}+j_{5}+j_{3}
\end{array}
\right.
\,,\qquad
\beta_{l}=
\left|
\begin{array}{l}
j_{1}+j_{2}+j_{4}+j_{5}\\
j_{1}+j_{3}+j_{4}+j_{6}\\
j_{2}+j_{3}+j_{5}+j_{6}
\end{array}
\right.\,.
$$
As studied in \cite{Freidel:1998ua}, one can study the expansion of the $q$-deformed 6j-symbol in terms of the cosmological constant $\Lambda$. At leading order around $\Lambda=0$, one recovers as expected the standard 6j-symbol, see fig.\ref{fig:6jq}. Then the first derivative is actually given by triple graspings and Casimir insertions\footnotemark{}:
\footnotetext{
Although a proof of this next-to-leading behavior to the $\{6j\}_{q}$-symbol is not presented in \cite{Freidel:1998ua}, private communication with Freidel revealed that the formula is based on the interpretation of the $\{6j\}_{q}$-symbol as the Drinfeld associator. Here, we don't exactly get the same numerical coefficients as in \cite{Freidel:1998ua}, due to different conventions for the operators.
}
\be
\pp_{\Lambda}\sixj{j_{1}}{j_{2}}{j_{3}}{j_{4}}{j_{5}}{j_{6}}_{q}\bigg{|}_{\Lambda\rightarrow0}\,=\,
+\f16\left(
\f14\sum_{i}j_{i}(j_{i}+1)
+
\hat{V}
\right)
\sixj{j_{1}}{j_{2}}{j_{3}}{j_{4}}{j_{5}}{j_{6}}\,,
\label{dq6j}
\ee
where the volume operator is defined as earlier as the sum over the four non-degenerate triple graspings, $\hat{V}=
(\hat{T}_{456}+\hat{T}_{234}+\hat{T}_{135}+\hat{T}_{126})/4$.

\begin{figure}[h!]

\begin{subfigure}[t]{.45\linewidth}
\includegraphics[scale=0.5]{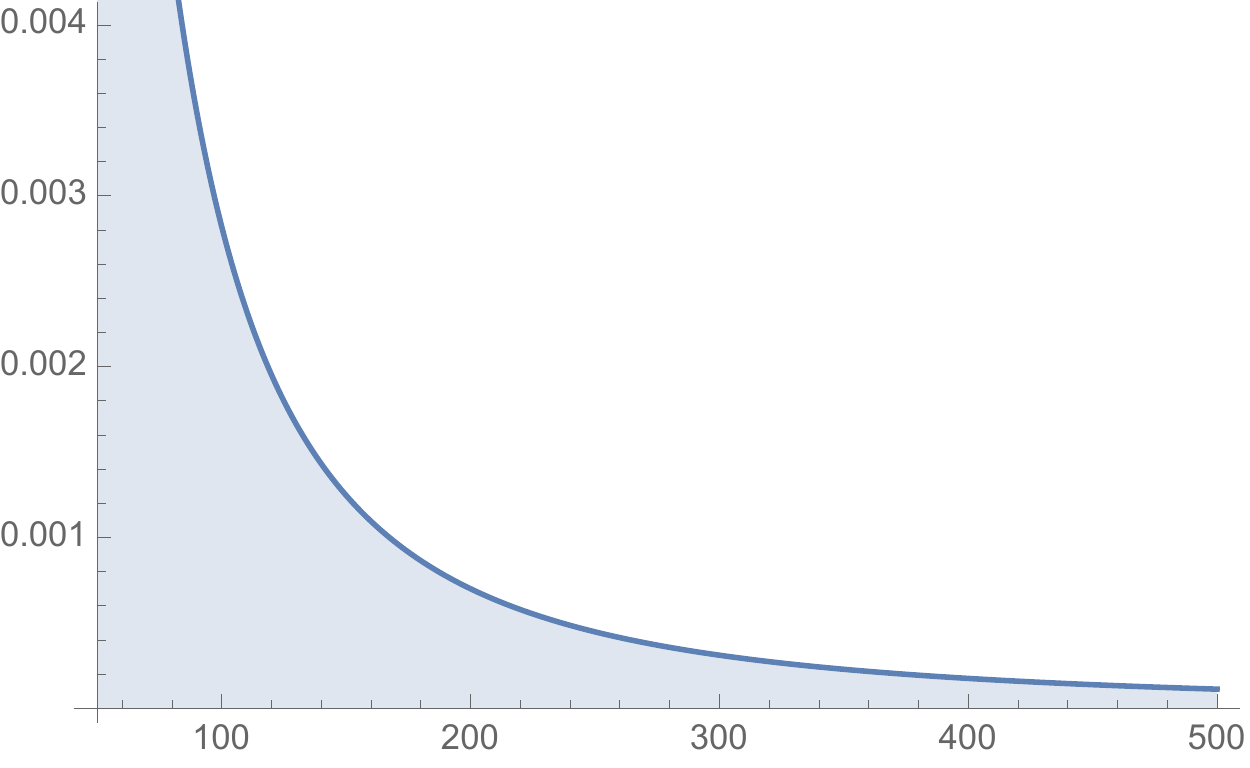}
\caption{$k$ ranges from 50 to 500: we see that the difference goes to 0 as $k$ goes to infinity.}
\end{subfigure}
\hspace{2mm}
\begin{subfigure}[t]{.45\linewidth}
\includegraphics[scale=0.5]{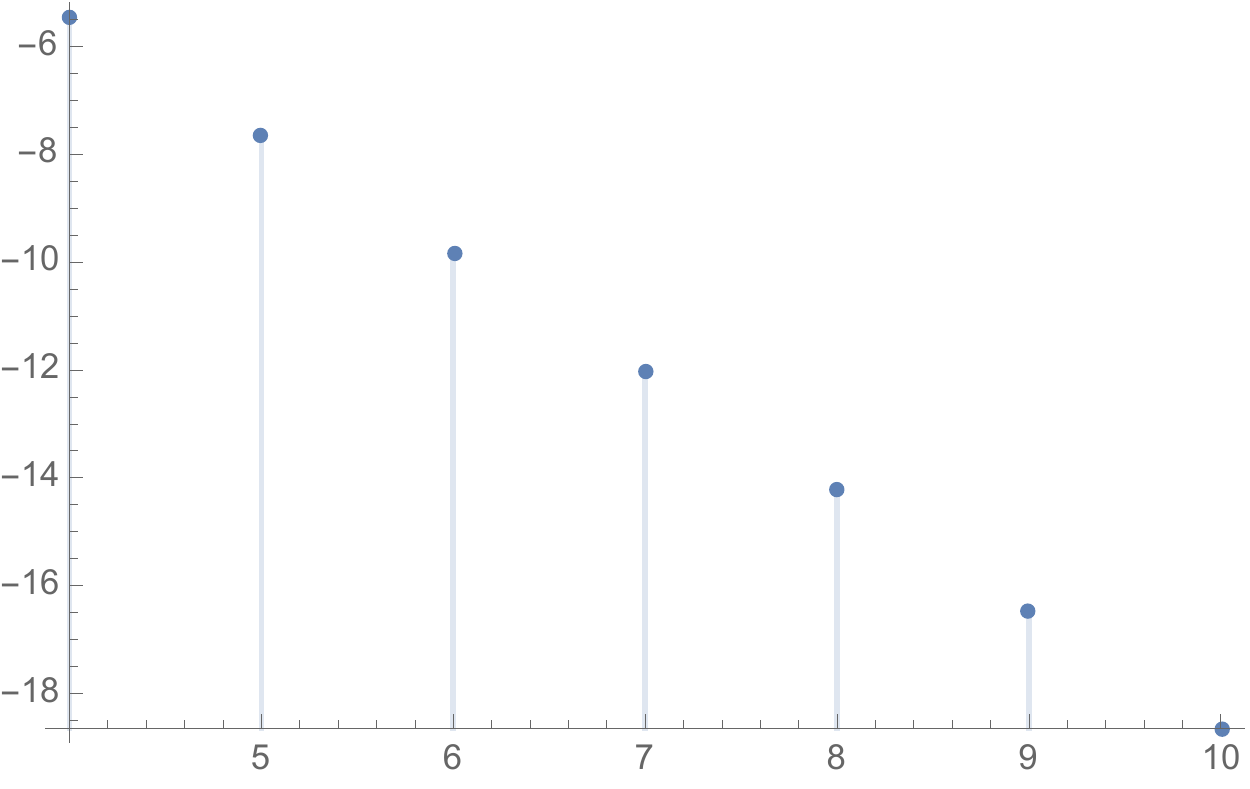}
\caption{Log-log plot with $k$ taking the values $3^{l}$ with $l$ ranging from 4 to 10: the slope is $-2$ showing that the difference between the $\{6j\}_{q}$ and the $\{6j\}$ decreases as $k^{-2}$.}
\end{subfigure} 

\caption{The difference  $\{6j\}_{q}-\{6j\}$ for $q=\exp({2i\pi}/k)$  for the homogeneous spin configuration $j_{i}=5$.}
\label{fig:dq6j}

\end{figure}

\begin{figure}[h!]


\begin{subfigure}[t]{.45\linewidth}
\includegraphics[scale=0.5]{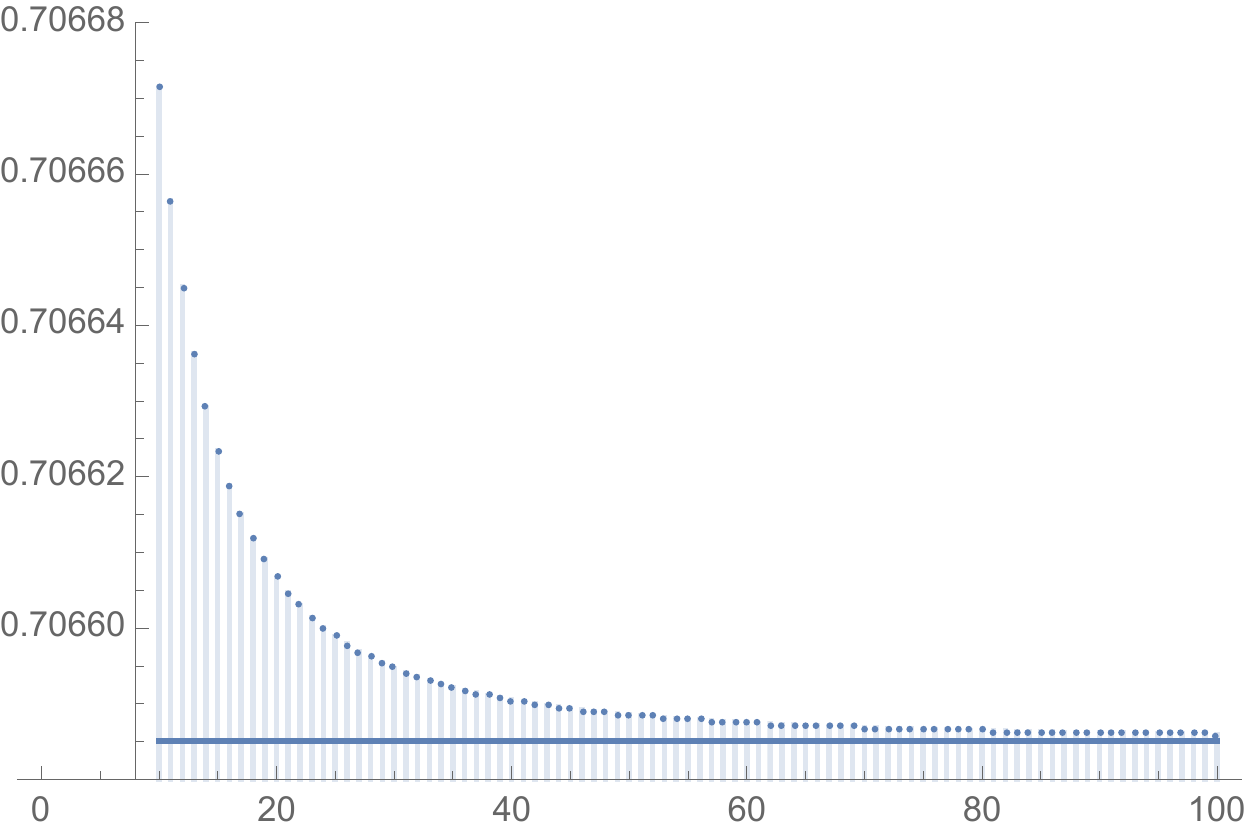}
\caption{The homogeneous spin configuration $j_{i}=5$: the triple grasping formula gives $2425/3432\sim 0.706585$.}
\end{subfigure}
\hspace{2mm}
\begin{subfigure}[t]{.45\linewidth}
\includegraphics[scale=0.5]{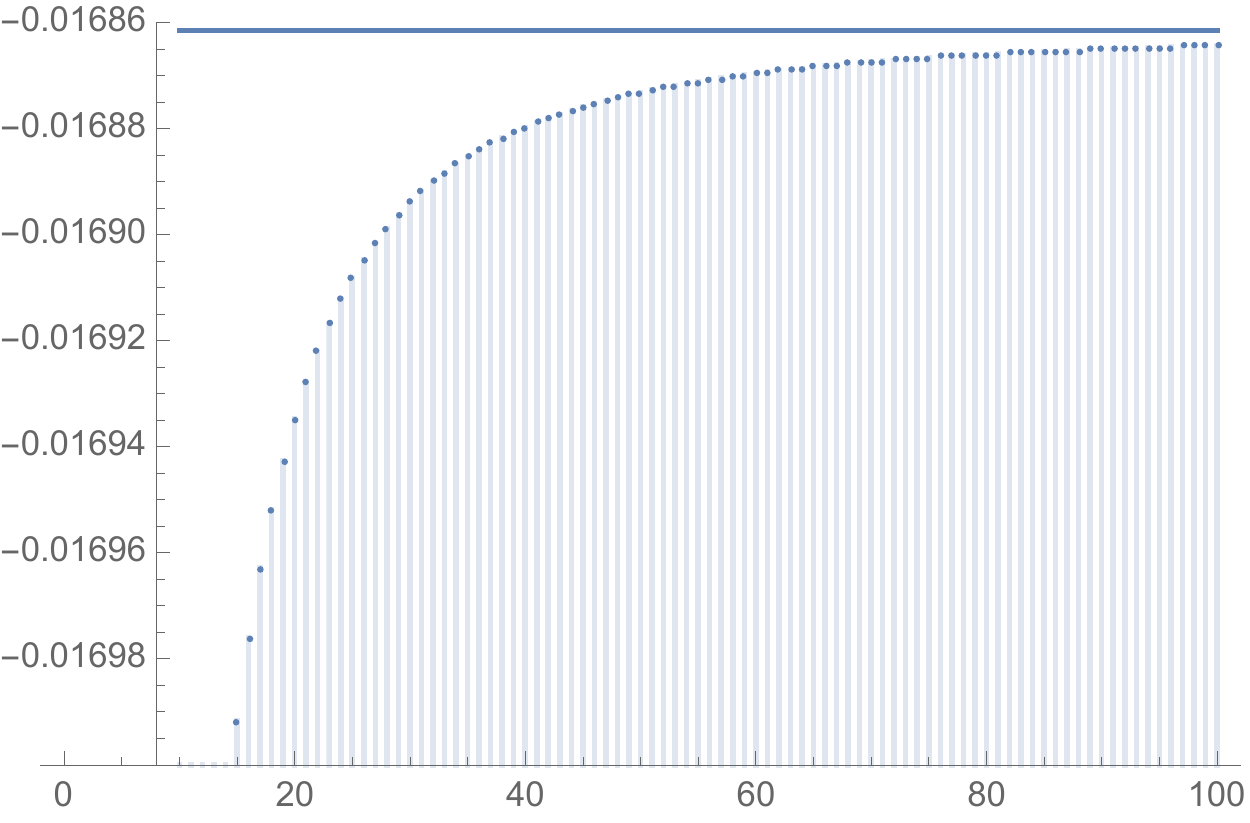}
\caption{The inhomogeneous spin configuration $j_{1}=6$, $j_{2} =5$, $j_{3}=4$, $j_{4}=7$, $j_{5}=5$, $j_{6}=6$: the triple grasping formula gives $-53\sqrt{19/255}/858\sim -0.0168615$.}
\end{subfigure} 

\caption{We check the first derivative formula \eqref{dq6j} for the $\{6j\}_{q}$-symbol for $q=\exp({2i\pi}/k)$ in the root-of-unity case. The discrete plot is the exact difference between $\{6j\}_{q}$ and $\{6j\}$, rescaled by $(k/2\pi)^{2}$, with $k$ ranging from $10^{3}$ to $10^{4}$. The horizontal axis gives the value of  $k/100$. It converges to the triple grasping formula for the first derivative \eqref{dq6j} and drawn as the continuous horizontal line. }
\label{fig:dq6j1}

\end{figure}

\begin{figure}[h!]

\begin{subfigure}[t]{.45\linewidth}
\includegraphics[scale=0.5]{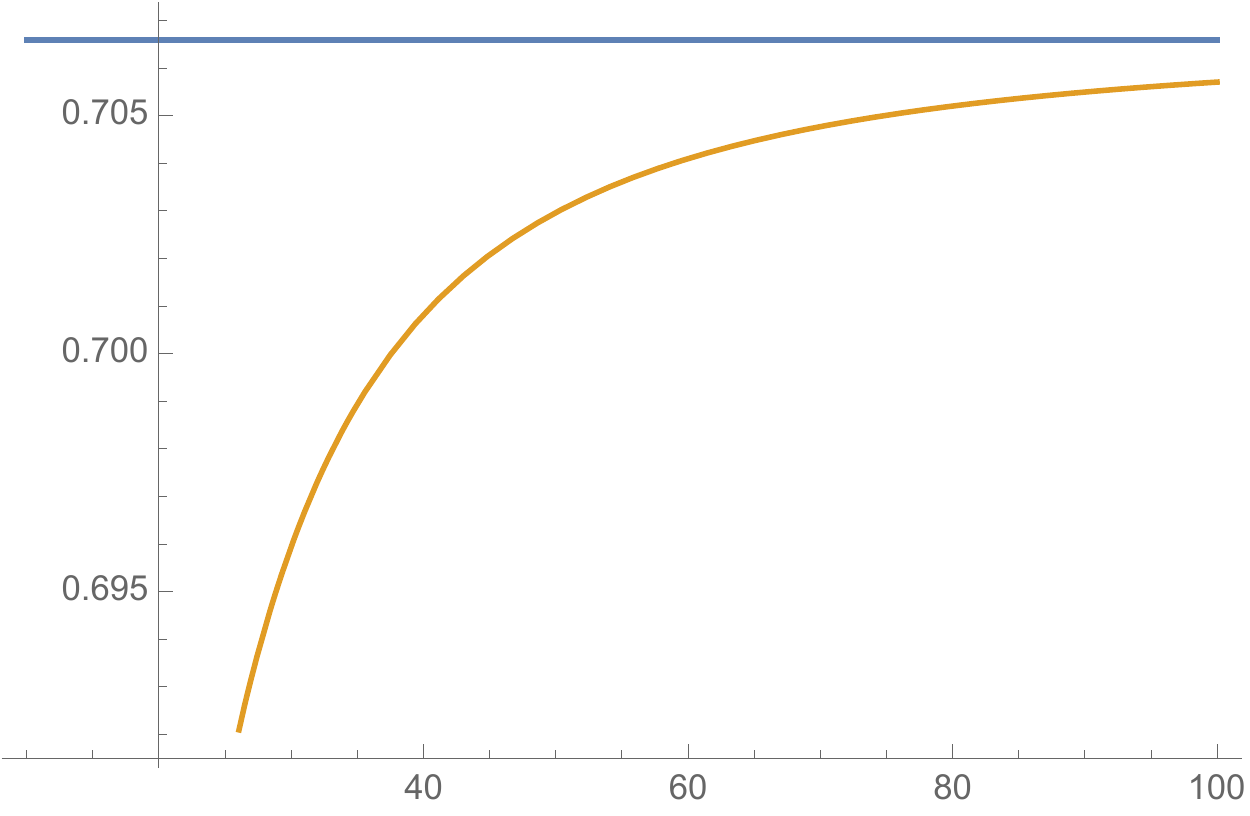}
\caption{The homogeneous spin configuration $j_{i}=5$: the triple grasping formula gives $2425/3432\sim 0.706585$.}
\end{subfigure}
\hspace{2mm}
\begin{subfigure}[t]{.45\linewidth}
\includegraphics[scale=0.5]{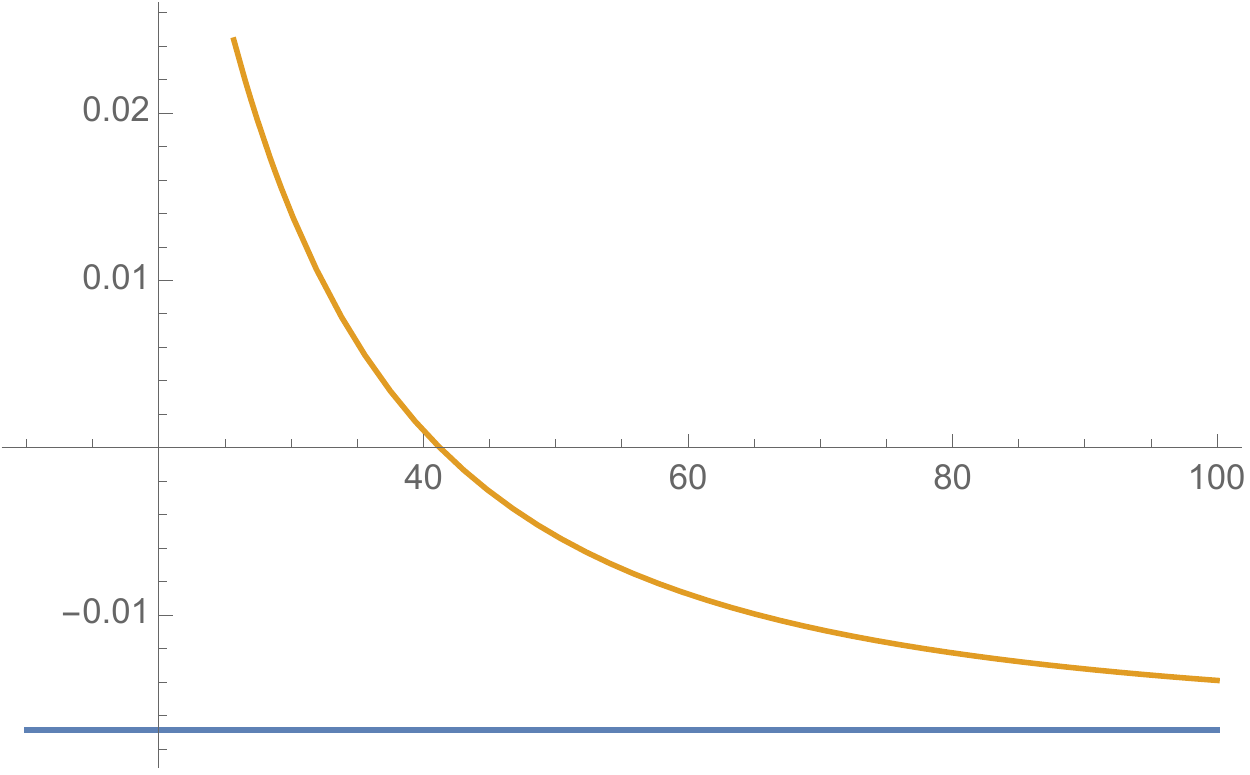}
\caption{The inhomogeneous spin configuration $j_{1}=6$, $j_{2} =5$, $j_{3}=4$, $j_{4}=7$, $j_{5}=5$, $j_{6}=6$: the triple grasping formula gives $-53\sqrt{19/255}/858\sim -0.0168615$.}
\end{subfigure} 

\caption{We check the first derivative formula \eqref{dq6j} for the $\{6j\}_{q}$-symbol for a real parameter $q=\exp(2\tau)$. We plot the exact difference between $\{6j\}_{q}$ and $\{6j\}$, rescaled by $1/(2\tau)^{2}$, with $\tau^{-1}$ ranging from $10$ to $10^{2}$. It converges to the triple grasping formula for the first derivative \eqref{dq6j}. }

\label{fig:dq6j2}

\end{figure}

This crucial formula establishes the link between the $q$-deformation of the $\SU(2)$ gauge group and the cosmological constant controlling the volume term for 3d (quantum) gravity.
We tested this formula numerically comparing the exact $\{6j\}_{q}$-symbol to its leading order approximation for $q\sim 1$ obtained by the action of the volume operator of the standard 6j-symbol\footnotemark{}:
\be
\sixj{j_{1}}{j_{2}}{j_{3}}{j_{4}}{j_{5}}{j_{6}}_{q=e^{\f{2i\pi}k}}
\quad
\underset{k\ge 1}{\sim}
\quad
\sixj{j_{1}}{j_{2}}{j_{3}}{j_{4}}{j_{5}}{j_{6}}
+\left(\f{2\pi}{k}\right)^{2}\,\f16\left(
\f14\sum_{i}j_{i}(j_{i}+1)
+
\hat{V}
\right)
\sixj{j_{1}}{j_{2}}{j_{3}}{j_{4}}{j_{5}}{j_{6}}\,,
\ee
\be
\sixj{j_{1}}{j_{2}}{j_{3}}{j_{4}}{j_{5}}{j_{6}}_{q=e^{2\tau}}
\quad
\underset{\tau\rightarrow 0}{\sim}
\quad
\sixj{j_{1}}{j_{2}}{j_{3}}{j_{4}}{j_{5}}{j_{6}}
-(2\tau)^{2}\,\f16\left(
\f14\sum_{i}j_{i}(j_{i}+1)
+
\hat{V}
\right)
\sixj{j_{1}}{j_{2}}{j_{3}}{j_{4}}{j_{5}}{j_{6}}\,.
\ee
\footnotetext{
For $q$ root of unity, one should remember that we do not a continuum of values for $q$ and $\Lambda$. However, this point is not so relevant here since we consider the limit $q\rightarrow 1$. One might have to be more careful when considering the variation of the $\{6j\}_{q}$-symbol around a non-trivial root of unity.
}
We provide in fig.\ref{fig:dq6j}, fig.\ref{fig:dq6j1} and in fig.\ref{fig:dq6j2} the results of numerical simulations of this leading order correction to the $q$-deformation of the 6j-symbol.

\subsection{The topological invariance of the $\{6j\}_{q}$-symbols order by order in $q$}

The representation theory of $\cU_{q}\SU(2)$ and combinatorics of the $q$-deformed Clebsch-Gordan coefficients are the same as for the standard $\SU(2)$ group and the Clebsch-Gordan coefficients. A direct consequence is that the $\{6j\}_{q}$-symbols satisfy the same invariance property under 3-2 and 4-1 Pachner moves. For instance, it satisfies the same identity \eqref{41move} as we derived earlier for the 6j-symbol:
\begin{multline} 
[d_{J}]_{q}\,\sixj{j_{1}}{j_{2}}{j_{3}}{j_{4}}{j_{5}}{j_{6}}_{q}=
\sum_{k_{4},k_{5},k_{6}}
(-1)^{\sum_{i}j_{i}+k_{4}+k_{5}+k_{6}+J}[d_{k_{4}}]_{q}[d_{k_{5}}]_{q}[d_{k_{6}}]_{q}\\
\times\sixj{k_{4}}{j_{4}}{J}{j_{6}}{k_{6}}{j_{2}}_{q}
\sixj{k_{5}}{j_{5}}{J}{j_{4}}{k_{4}}{j_{3}}_{q}
\sixj{k_{6}}{j_{6}}{J}{j_{5}}{k_{5}}{j_{1}}_{q}
\sixj{j_{1}}{j_{2}}{j_{3}}{k_{4}}{k_{5}}{k_{6}}_{q}\,.
\label{q41move}
\end{multline}
We checked that this identity holds for both cases $q\in\U(1)$ and $q\in\R_{+}$. It leads to the topological invariance of the Turaev-Viro amplitude \cite{Turaev:1992hq}, constructed just like the Ponzano-Regge model but associating a $\{6j\}_{q}$-symbol instead of the 6j-symbol to each tetrahedron of the 3d triangulation.

At leading order, in the classical limit $q\rightarrow 1$, the identity above leads back to the 4-1 Pachner move identity  \eqref{41move} for the 6j-symbol. On the other hand, in light of the derivative formula \eqref{dq6j} for the $\{6j\}_{q}$-symbol, the next-to-leading order of the $q$-deformed 4-1 Pachner move identity, expanding the formula at first order in the deformation parameter $q$, involves the insertion of Casimirs  (double graspings) and volume operators (triple graspings). We show here that the Pachner move invariance of the $\{6j\}_{q}$-symbol at first order in $q$ follows from the propagation of the length and volume observables under 4-1 Pachner moves as studied in the previous section.

\medskip

So our goal here is to derive the 4-1 Pachner move \eqref{q41move} for the $\{6j\}_{q}$-symbol at first order in $q$ from our improved identities of the topological invariance of the classical 6j-symbols with length and volume insertions. The first order in $q$ of the left hand side
of \eqref{q41move} simply involves the first derivatives of the $q$-dimension and of the $\{6j\}_{q}$-symbol. We gather these two contributions from \eqref{qdim} and \eqref{dq6j}:
\be
6\,\pp_{\Lambda}\,\textrm{l.h.s. of \eqref{q41move}}\,\bigg{|}_{\Lambda=0}
\,=\,
d_{J}\,\left(
\f14\sum_{i}j_{i}(j_{i}+1)
+
\hat{V}
\right)
\sixj{j_{1}}{j_{2}}{j_{3}}{j_{4}}{j_{5}}{j_{6}}
-
d_{J}\,\cC_{J}\,\sixj{j_{1}}{j_{2}}{j_{3}}{j_{4}}{j_{5}}{j_{6}}
\,.
\label{lhs}
\ee
The right hand side contains many more terms, with $\Phi=(\sum_{i}j_{i})+(k_{4}+k_{5}+k_{6})+J$:
\beq
6\,\pp_{\Lambda}\,\textrm{r.h.s. of \eqref{q41move}}\,\bigg{|}_{\Lambda=0}=
&&
\sum_{k_{4},k_{5},k_{6}}
(-1)^{\Phi}d_{k_{4}}d_{k_{5}}d_{k_{6}}\,\Bigg{[}
-(\cC_{k_{4}}+\cC_{k_{5}}+\cC_{k_{6}})
\{6j\}\{6j\}\{6j\}\{6j\}
\label{rhs}
\\
&&+\left(
\f14(\cC_{k_{4}}+\cC_{j_{4}}+\cC_{J}+\cC_{j_{6}}+\cC_{k_{6}}+\cC_{j_{2}})
+
\hat{V}
\right)\sixj{k_{4}}{j_{4}}{J}{j_{6}}{k_{6}}{j_{2}}
\{6j\}\{6j\}\{6j\}
\nn\\
&&+
\{6j\}\left(
\f14(\cC_{k_{5}}+\cC_{j_{5}}+\cC_{J}+\cC_{j_{4}}+\cC_{k_{4}}+\cC_{j_{3}})
+
\hat{V}
\right)\sixj{k_{5}}{j_{5}}{J}{j_{4}}{k_{4}}{j_{3}}
\{6j\}\{6j\}
\nn\\
&&+\{6j\}\{6j\}\left(
\f14(\cC_{k_{6}}+\cC_{j_{6}}+\cC_{J}+\cC_{j_{5}}+\cC_{k_{5}}+\cC_{j_{1}})
+
\hat{V}
\right)\sixj{k_{6}}{j_{6}}{J}{j_{5}}{k_{5}}{j_{1}}
\{6j\}
\nn\\
&&+\{6j\}\{6j\}\{6j\}\left(
\f14(\cC_{j_{1}}+\cC_{j_{2}}+\cC_{j_{3}}+\cC_{k_{4}}+\cC_{k_{5}}+\cC_{k_{6}})
+
\hat{V}
\right)\sixj{j_{1}}{j_{2}}{j_{3}}{k_{4}}{k_{5}}{k_{6}}\Bigg{]}\,.
\nn
\eeq
Equating the left hand and right hand sides above in \eqref{lhs} and \eqref{rhs}, we realize that the first order in $q$ of the 4-1 Pachner move invariance of the $\{6j\}_{q}$-symbol has a simple geometrical interpretation: the sum of the volume of the four tetrahedra of the finer triangulation equates the total volume of the single tetrahedron of the coarse triangulation in the 4-1 Pachner move, as drawn earlier on fig.\ref{tent}, up to sub-leading corrections (the Casimirs are in $j^{2}$ while the volumes are in $j^{3}$).

Now let us prove the exact equality between the left and right hand side derivatives given above.
We already know how to deal explicitly with the Casimir insertions and the volume operator (triple graspings) acting on the fourth and last 6j-symbol factor, using the improved identities \eqref{pachner41length} and \eqref{pachner41volume}. They propagate through the 4-1 Pachner move up to $\cC_{J}$ terms, which we can keep track of explicitly. Up to an overall $d_{J}$-factor, a Casimir term $\cC_{k_{i}}$ will sum to the corresponding $\cC_{j_{i}}$ plus a $\cC_{J}$ contribution, while the volume $\hat{V}$ acting on $\{j's,k's\}$ propagates  through the sums with a $-\f14\cC_{J}$ correction.
The three other volume operator insertions actually satisfy a much simpler formula:
\begin{multline} 
-{d_{J}}\,(\cC_{J}+\cC_{j_{2}})\,\sixj{j_{1}}{j_{2}}{j_{3}}{j_{4}}{j_{5}}{j_{6}}
=
\sum_{k_{4},k_{5},k_{6}}
(-1)^{\sum_{i}j_{i}+k_{4}+k_{5}+k_{6}+J}d_{k_{4}}d_{k_{5}}d_{k_{6}}\,
\\
\times
\left[\hat{V}\sixj{k_{4}}{j_{4}}{J}{j_{6}}{k_{6}}{j_{2}}\right]\,
\sixj{k_{5}}{j_{5}}{J}{j_{4}}{k_{4}}{j_{3}}
\sixj{k_{6}}{j_{6}}{J}{j_{5}}{k_{5}}{j_{1}}\,
\sixj{j_{1}}{j_{2}}{j_{3}}{k_{4}}{k_{5}}{k_{6}}\,,
\label{pachner41volA}
\end{multline}
and similarly for the two other 6j-symbols involving $j_{3}$ and $j_{1}$.
This means that the volume of these three tetrahedra vanish in average during the 4-1 Pachner move and we simply get sub-leading quantum corrections given by some quadratic Casimir operators.
At the end of the day, we simply count the number of occurrence of each Casimir $\cC_{j_{i}}$ and $\cC_{J}$ and find that the left hand side and right hand side do match.
This proves the topological invariance of the $\{6j\}_{q}$-symbol at first order in the deformation parameter $q$ from the identities on the classical 6j-symbol giving the behavior of the lengths and volumes under coarse-graining by 4-1 Pachner moves.

\medskip

To summarize, one can expand the 4-1 Pachner move invariance of the $\{6j\}_{q}$-symbol order by order in the deformation parameter $q$. At the zeroth order, we get the standard invariance of the 6j-symbol under the 4-1 Pachner move, which is equivalent to the Biedenharn-Elliott identity.
Here we have seen that, at first order, the $q$-deformation amounts to insertions of length and volume operators on the 6j-symbol, in such a way to preserve the invariance under the 4-1 Pachner move.
We expect this to be true at all orders of the expansion in $q$, and it would surely be  enlightening to derive the precise ordering of the grasping operators at each order of the expansion of the $\{6j\}_{q}$-symbol in order to understand explicit the $q$-deformation of the Ponzano-Regge amplitude and its geometrical interpretation at all orders in $q$.

\section*{Conclusion}

In this paper, we looked in great detail at the invariance of the Ponzano-Regge model under Pachner moves, which imply the topological invariance of the Ponzano-Regge amplitudes and validates its interpretation as the discrete path integral for 3d Euclidean quantum gravity. We focused in particular on the 4-1 Pachner move, which corresponds to mapping a single tetrahedron to four tetrahedra by introducing an extra bulk vertex and vice-versa. This move can be interpreted as the basic coarse-graining move for 3d triangulations. Mathematically, the invariance of the Ponzano-Regge amplitude under the 4-1 Pachner move is equivalent to the Biedenharn-Elliott identity for the 6j-symbol. Here we produced extensions of this identity accounting for the behavior under coarse-graining of length and volume observables and we showed how these new identities are related to the Biedenharn-Elliott identity for the $q$-deformed $\{6j\}_{q}$-symbol. 

First, we recalled how the 4-1 Pachner move invariance of the 6j-symbol can be generated by acting with a holonomy operator on the tetrahedral spin network and evaluating the resulting state against the flat state, as originally showed in \cite{Bonzom:2009zd}. This is especially relevant since the holonomy operators are the Hamiltonian constraints for 3d quantum gravity in the framework of loop quantum gravity and spinfoam models. By considering higher derivative of the flat state as suggested in \cite{Charles:2016xwc}, this technique allowed to derive extended Biedenharn-Elliott identities for the 6j-symbol with length and volume insertions. These insertions are realized as double and triple graspings on the tetrahedral spin network and we used the explicit formula worked out in \cite{Hackett:2006gp}. Our new extended Biedenharn-Elliott identities show how lengths and volumes transform under coarse-graining: they are almost invariant and one can control explicitly the correction terms.

Second, we considered the $q$-deformation of the Ponzano-Regge model to the Turaev-Viro model. The $q$-deformation is supposed to account for the introduction of a non-vanishing cosmological constant $\Lambda$ in the 3d gravity path integral. In Euclidean space-time signature (+++), the spherical case with a positive cosmological constant $\Lambda>0$ corresponds to $q=\exp\,i\sqrt{\Lambda}$ root-of-unity, while the hyperbolic case with a negative cosmological constant $\Lambda<0$ corresponds to a real deformation parameter $q=\exp\,\sqrt{-\Lambda}$.
This is mostly on the basis that the large spin asymptotics of the $\{6j\}_{q}$-symbol adds a volume term to the Regge action, although recent work have confirmed this relation in a canonical framework (space/time splitting) for both $\Lambda>0$ \cite{Noui:2011im,Noui:2011aa,Pranzetti:2014xva} and $\Lambda<0$ \cite{Dupuis:2013haa,Bonzom:2014bua,Bonzom:2014wva}.
In the present context of coarse-graining and 4-1 Pachner moves, the Biedenharn-Elliott identity for the $\{6j\}_{q}$-symbol implies the topological invariance of the Turaev-Viro amplitudes. We looked at the expansion of this 4-1 Pachner move invariance of the $\{6j\}_{q}$-symbol order by order in the deformation parameter $q$.
In the classical limit $q\rightarrow1$,  the $\{6j\}_{q}$-symbol leads back to the standard 6j-symbol. Then the next order in $q$, given by the first derivative with respect to $q$, is given by the volume operator acting on the 6j-symbol up to some Casimir terms which are subleading. This volume operator is realized though triple graspings. Our formula confirms the result of \cite{Freidel:1998ua}, up to some numerical factors due to different sign and normalization conventions. This allows us to prove the 4-1 Pachner move invariance of the $\{6j\}_{q}$-symbol at first order $q$ from our extended Biedenharn-Elliott identities for the 6j-symbol with length and volume insertions. It also provides it with a simple geometrical interpretation: the sum of the four tetrahedron volumes equal the volume of the full tetrahedron in a 4-1 Pachner move and this holds as an operator identity at the quantum level.

\medskip

Our analysis can be considered as an extra step towards validating the $q$-deformation of quantum gravity amplitudes to account for a cosmological constant.
Looking at the variations in $q$ of the $\{6j\}_{q}$-symbol and its 4-1 Pachner move invariance allowed to understand the $q$-deformation at leading order in terms of the volume operator. Pushing this further should allow to better  understand the geometrical interpretation and the quantum gravity interpretation of the ``classical'' limit $q\rightarrow 1$ of the Turaev-Viro topological invariant.
For instance, can we compute explicitly all the derivatives with respect to $q$ of the $\{6j\}_{q}$-symbol at $q=1$? Could the $\{6j\}_{q}$-symbol be obtained from the standard 6j-symbol by acting with an exponentiated volume operator\footnotemark{}?
\footnotetext{To define an exponentiated volume operator, one should take special care of the ordering of the triple grasping operators. This issue reflects the ambiguities in the $q$-deformation of the amplitudes due the non-trivial braiding for $q\ne 1$.}
Investigating the higher orders of the expansion series  for the $\{6j\}_{q}$-symbol in terms of the cosmological constant would lead to a true understanding of the continuous deformation by the parameter $q$. A goal could be to derive a differential equation in $q$ for the  $\{6j\}_{q}$-symbol. A seemingly promising point of view is to use the formula expressing it as the Drinfled associator and its relation to a WZW conformal field, as used in \cite{Freidel:1998ua} and hinted to in \cite{Roberts:2002}\footnotemark{}.
\footnotetext{
Quoting  \cite{Roberts:2002}: ``\textit{If we view the moduli space as $\C\setminus \{0, 1\}$ then we seek the holonomy along the unit interval from 0 to 1. Now asymptotically the connection we are examining becomes the Knizhnik-Zamolodchikov connection, and this holonomy is nothing more than the Drinfeld associator. (See Bakalov and Kirillov, for example.) This is the geometric explanation for the equivalence of the 6j-symbol and associator pointed out recently by Bar-Natan and Thurston. Of course, one could try to compute a nice tetrahedrally symmetric formula for the associator
hoping that the asymptotic formula for the 6j-symbol would follow: this would
be a completely alternative approach to the asymptotic problem.}''
}

This line of research underlines the importance to study the flow of the quantum gravity amplitudes with respect to the deformation parameter $q$. The $q$-deformation is usually introduced algebraically in terms of quantum groups, categories and fusion algebra. Looking at the flow in $q$ starting at its classical limit $q\rightarrow 1$ allows a more geometrical point of view, as here when we express the $\{6j\}_{q}$-symbol at first order in $q$ in terms of the volume operator acting on the 6j-symbol. Beyond leading to a better geometrical understanding of the $q$-deformed amplitude, analyzing this  $q$-flow will become essential in 4d quantum gravity when we expect the cosmological constant to flow under the renormalization group.

Finally, we conclude  with two possible direct extensions of the present work. On the one hand, we should generalize our analysis to the Lorentzian case and apply the same methods to the 6j-symbol for the $\SU(1,1)$ Lie group. On the other hand, it would be very interesting to compare the first order expansion in $q$ of the Turaev-Viro amplitude, as studied here, to the way to introduce the cosmological constant in the Regge calculus path integral for 3d gravity.

\section*{Acknowledgement}

I am especially grateful to Laurent Freidel, for his explanations on the $q$-deformed 6j-symbol and its definition as the Drinfeld associator, and to Simone Speziale for his helpful comments and a careful reading of this paper.
I would also like to thank John Barrett, Christoph Charles, Bianca Dittrich, Aldo Riello and Marc Geiller for  their interest  on this project.

All numerical simulations and plots were realized using Mathematica 10. Please contact me by email for my mathematica files.

\appendix

\section{Wigner 3j-symbol}
\label{def3j}

In the Ponzano-Regge mode, each triangle is made of three spins, living on its three edges, and we attach to it the unique normalized intertwiner between the three corresponding representations. Writing $\cV^j$ for the $(2j+1)$-dimensional Hilbert space carrying the spin-$j$ representation, a intertwiner between three spins $j_{1,2,3}$ is a $\SU(2)$-invariant linear map from the tensor product $\cV^{j_{1}}\otimes\cV^{j_{2}}\otimes\cV^{j_{3}}$ to $\C$. There exists a non-trivial  intertwiner only if the three spins satisfy triangular inequalities, $|j_{2}-j_{3}|\le j_{1}\le(j_{2}+j_{3})$ and so on. It is then uniquely given by the normalized 3j-Wigner symbol (or equivalently the Clebsch-Gordan coefficients):
\be
C^{j_{1}j_{2}j_{3}}\,:\,\cV^{j_{1}}\otimes\cV^{j_{2}}\otimes\cV^{j_{3}}\longmapsto\C\,,
\qquad
\forall g\in\SU(2)\,,\,\,
C^{j_{1}j_{2}j_{3}}\circ g
=
C^{j_{1}j_{2}j_{3}}\circ (D^{j_{1}}(g)\otimes D^{j_{2}}(g) \otimes D^{j_{3}}(g))
=
C^{j_{1}j_{2}j_{3}}\,,
\ee
\be
C^{j_{1}j_{2}j_{3}}_{m_{1}m_{2}m_{3}}
=
C^{j_{1}j_{2}j_{3}}\,|j_{1}m_{1}\ra\otimes|j_{2}m_{2}\ra\otimes|j_{3}m_{3}\ra\,,
\qquad
\tr \,C^{j_{1}j_{2}j_{3}\,\dagger}\,C^{j_{1}j_{2}j_{3}}
=
\sum_{\{m_{i}\}}|C^{j_{1}j_{2}j_{3}}_{m_{1}m_{2}m_{3}}|^2
=
1\,.
\ee

\section{The triple grasping and the volume of the tetrahedron}
\label{V3}

\begin{figure}[h!]


\begin{subfigure}[t]{.45\linewidth}
\includegraphics[scale=0.5]{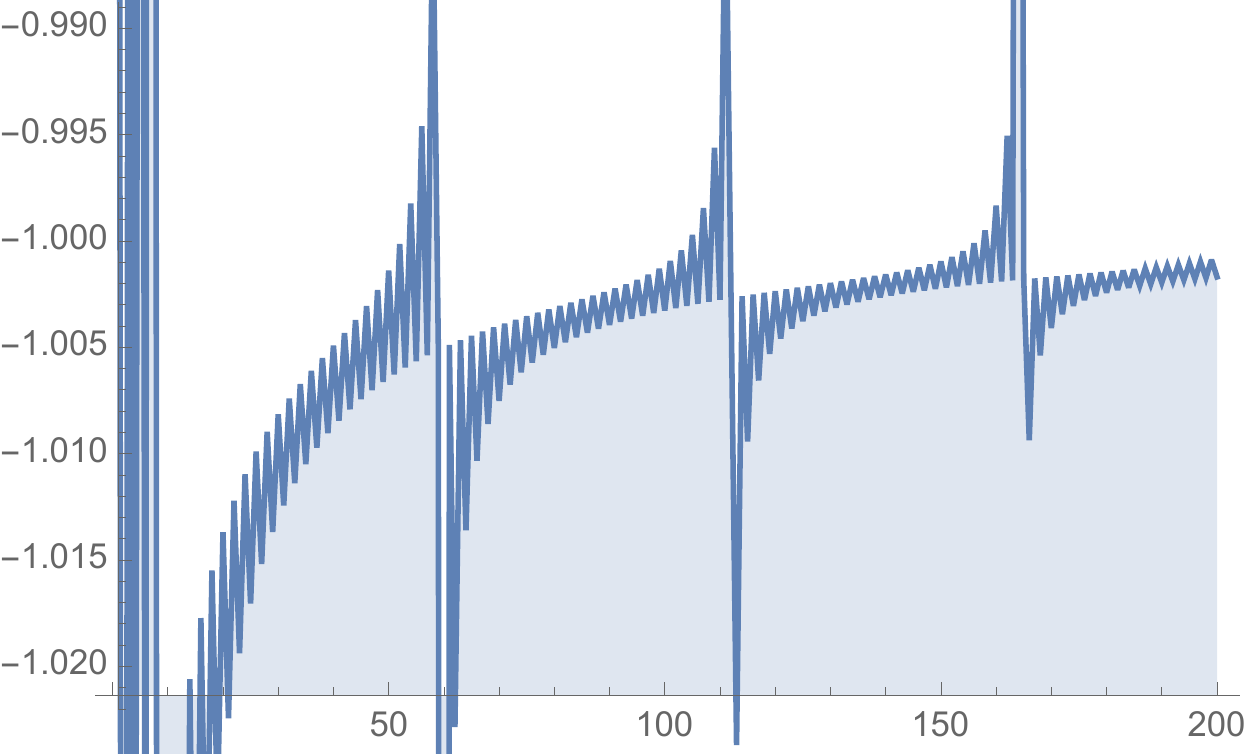}
\caption{Homogeneous spin configuration: the spins are all $10j$ with the parameter $j$ ranging from 1 to 200.}
\end{subfigure}
\hspace{2mm}
\begin{subfigure}[t]{.45\linewidth}
\includegraphics[scale=0.5]{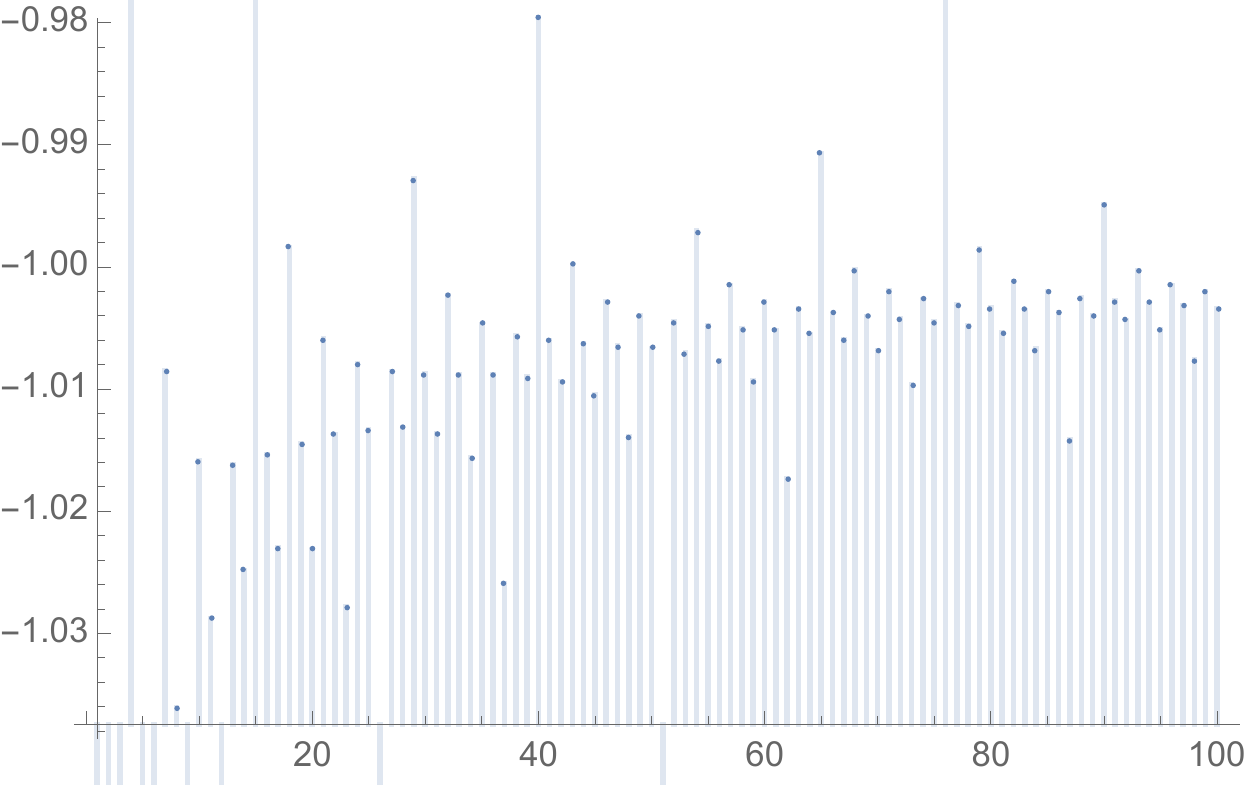}
\caption{Inhomogeneous spin configuration: the spins are $3j\times(6,5,4,7,5,6)$ with $j$ ranging from 1 to 100.}
\end{subfigure} 
\\
\begin{subfigure}[t]{.45\linewidth}
\includegraphics[scale=0.5]{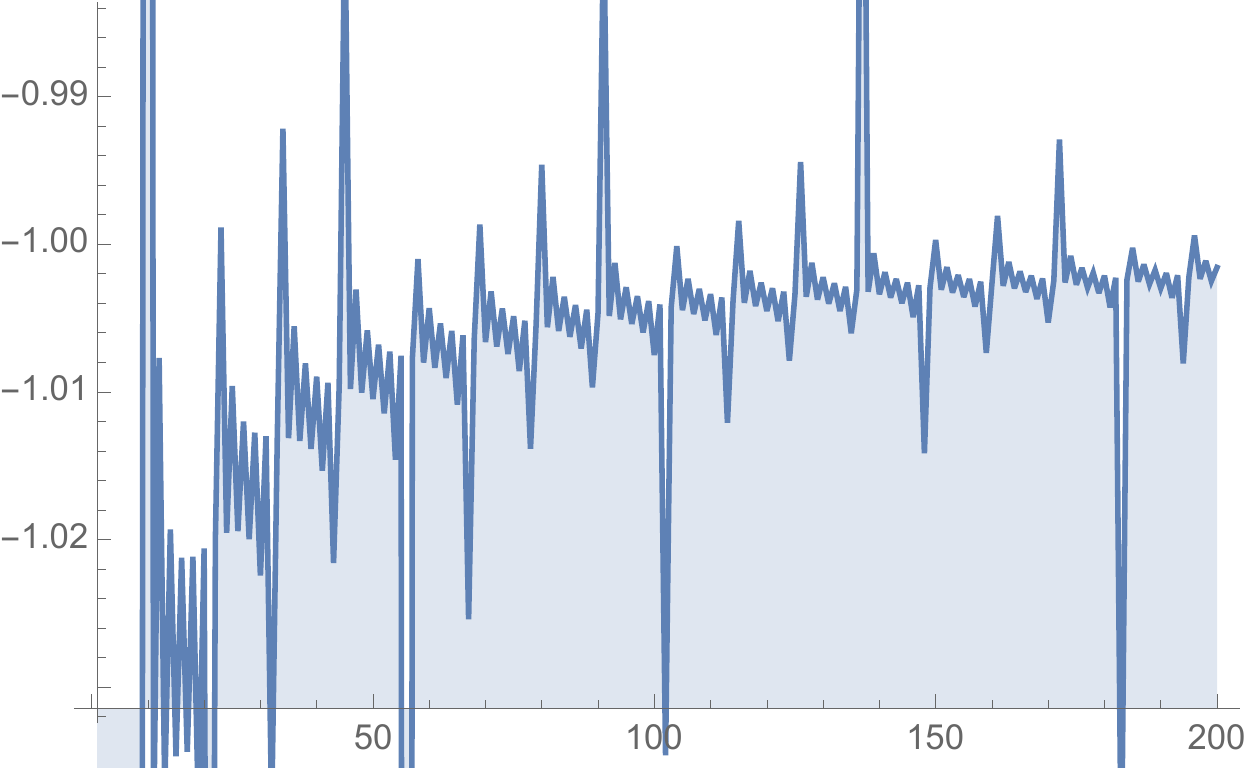}
\caption{Homogeneous spin configuration: the spins are all $7j$ with the parameter $j$ ranging from 1 to 200.}
\end{subfigure}
\hspace{2mm}
\begin{subfigure}[t]{.45\linewidth}
\includegraphics[scale=0.5]{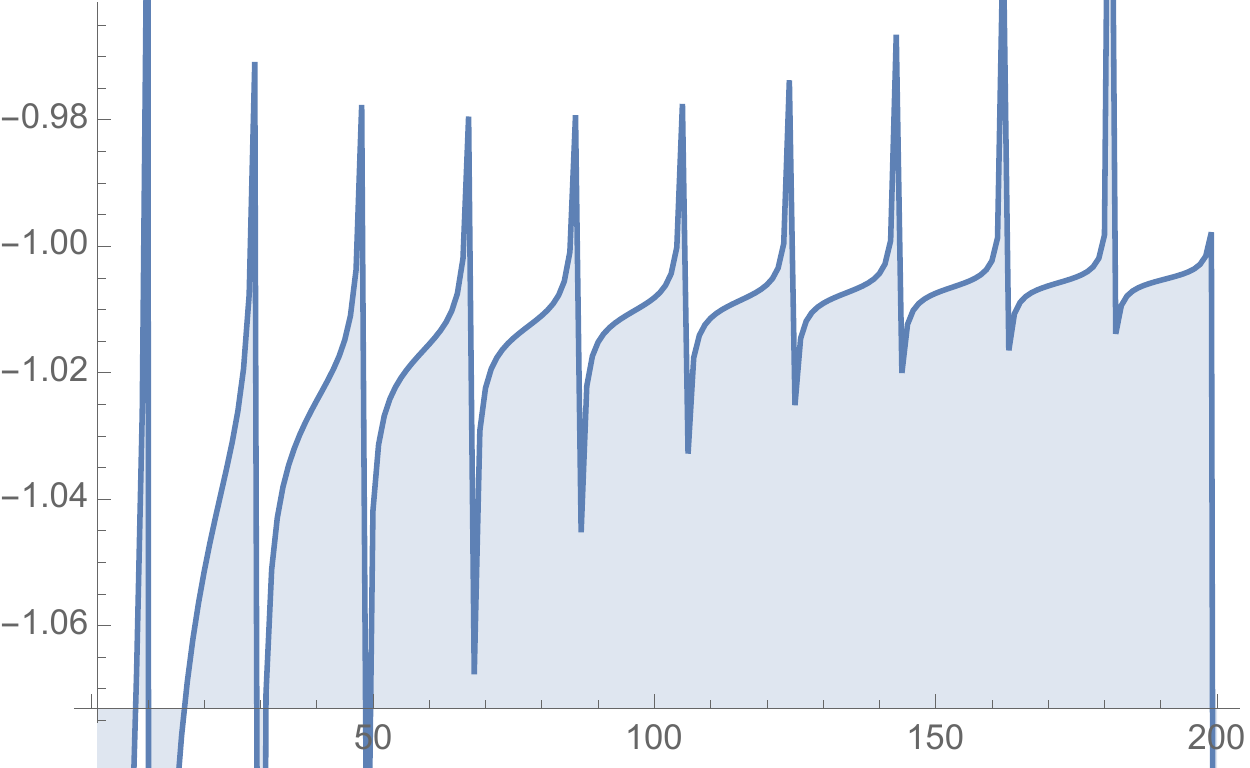}
\caption{Homogeneous spin configuration: the spins are all $3j$ with the parameter $j$ ranging from 1 to 200.}
\end{subfigure} 
\caption{The large spin limit of the squared triple grasping operator: we recover minus the volume of the classical tetrahedron. We nevertheless have wide oscillations due to the natural oscillatory behavior of the 6j-symbol.}
\label{fig:triplegrasping}

\end{figure}

\begin{figure}[h!]


\begin{subfigure}[t]{.45\linewidth}
\includegraphics[scale=0.5]{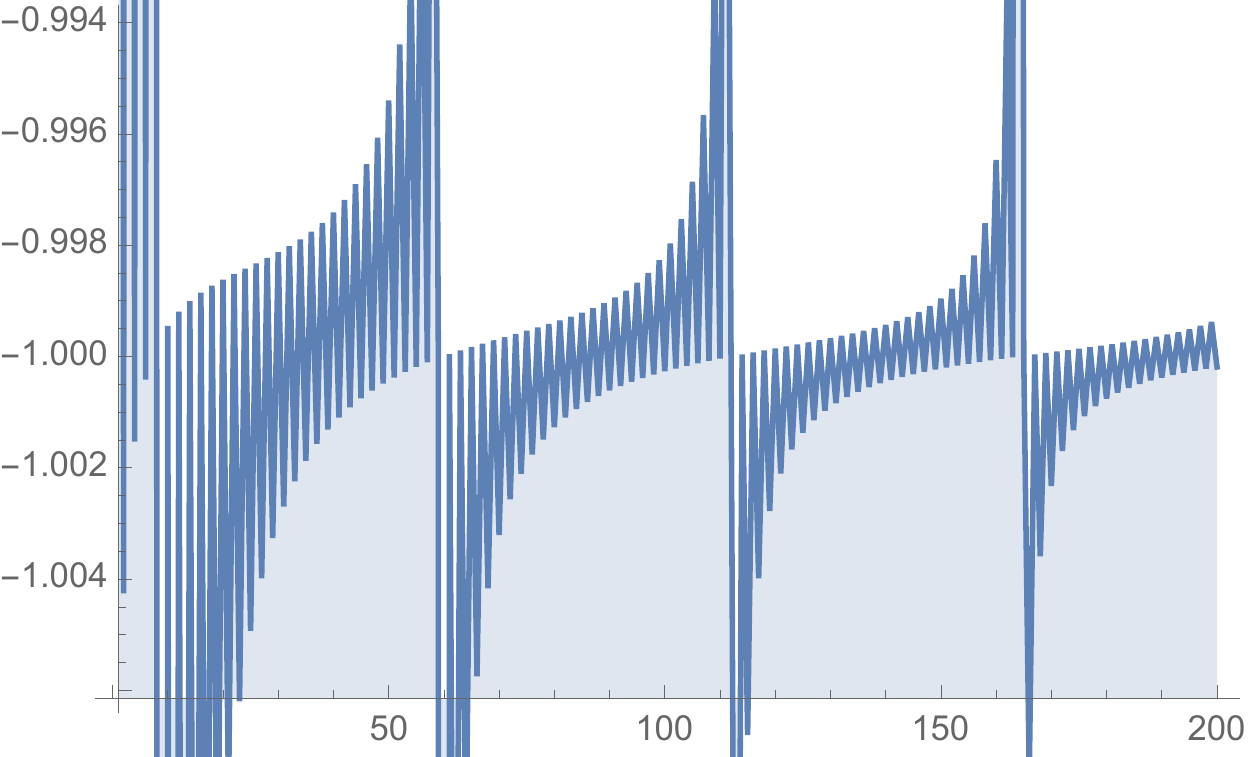}
\caption{Homogeneous spin configuration: the spins are all $10j$ with the parameter $j$ ranging from 1 to 200.}
\end{subfigure}
\hspace{2mm}
\begin{subfigure}[t]{.45\linewidth}
\includegraphics[scale=0.5]{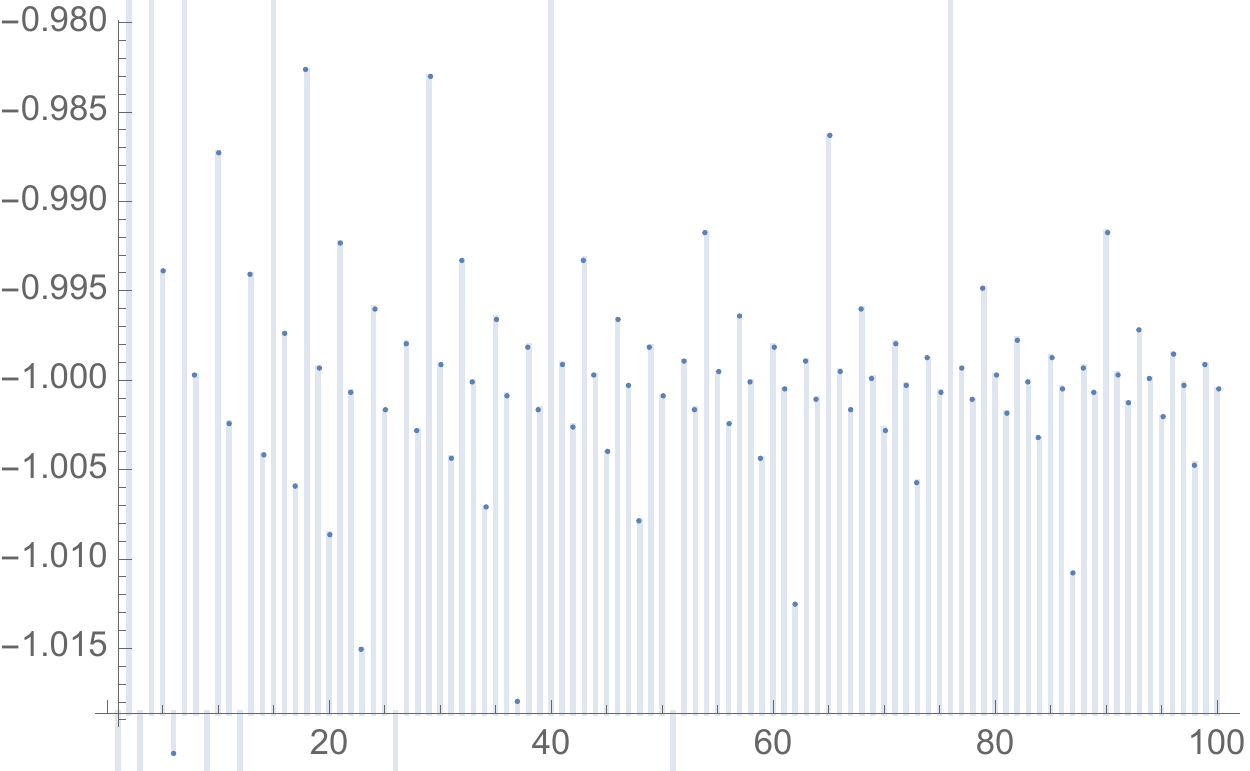}
\caption{Inhomogeneous spin configuration: the spins are $3j\times(6,5,4,7,5,6)$ with $j$ ranging from 1 to 100.}
\end{subfigure} 

\caption{We improve the fit of the (squared) triple grasping with the (squared) tetrahedron volume by shifting the edge length to  $j_{i}+\f12$.}
\label{fig:shiftedvolume}

\end{figure}

The triple grasping operator $\hat{T}_{456}=i\vJ^{(4)}\cdot(\vJ^{(5)}\w\vJ^{(6)})$ is the quantization of the classical volume observable for the tetrahedron up to a factor $3!=6$. This numerical factor is a simple symmetry factor, which can be thought of as the ratio between the volume of a cube and a tetrahedron: the cube contains 6 tetrahedra.
The squared volume of the tetrahedron is given as a polynomial of the edge lengths $j_{i}$:
\begin{multline}
U
\,=\,
\f1{12^{2}}\bigg{[}
(j_{1}^{2}+j_{2}^{2}+j_{3}^{2}+j_{4}^{2}+j_{5}^{2}+j_{6}^{2})(j_{1}^{2}j_{4}^{2}+j_{2}^{2}j_{5}^{2}+j_{3}^{2}j_{6}^{2})
-2j_{1}^{2}j_{4}^{2}(j_{1}^{2}+j_{4}^{2})-2j_{2}^{2}j_{5}^{2}(j_{2}^{2}+j_{5}^{2})-2j_{3}^{2}j_{6}^{2}(j_{3}^{2}+j_{6}^{2})
\\
-\f12(j_{1}^{2}+j_{4}^{2})(j_{2}^{2}+j_{5}^{2})(j_{3}^{2}+j_{6}^{2})
+\f12(j_{1}^{2}-j_{4}^{2})(j_{2}^{2}-j_{5}^{2})(j_{3}^{2}-j_{6}^{2})\,,
\bigg{]}
\end{multline}
which reduces to $U=j^{6}/72$ as expected in the equilateral case.
We recall the action of the triple grasping operator $\hat{T}_{456}$ on the 6j-symbol derived in \cite{Hackett:2006gp}:
\beq
\hat{T}_{456}\sixj{j_{1}}{j_{2}}{j_{3}}{j_{4}}{j_{5}}{j_{6}}
&=&
\alpha_{-}\sixj{j_{1}}{j_{2}}{j_{3}}{j_{4}-1}{j_{5}}{j_{6}}
+\alpha_{0}\sixj{j_{1}}{j_{2}}{j_{3}}{j_{4}}{j_{5}}{j_{6}}
+\alpha_{+}\sixj{j_{1}}{j_{2}}{j_{3}}{j_{4}+1}{j_{5}}{j_{6}}\,,
\eeq
where the $\alpha$-coefficients are defined as:
\be
\alpha_{-}
\,=\,
\f{j_{4}(j_{4}+1)}{4(2j_{4}+1)}\alpha(j_{1},j_{2},j_{3},j_{4},j_{5},j_{6})\,,
\qquad
\alpha_{+}
\,=\,
\f{j_{4}(j_{4}+1)}{4(2j_{4}+1)}\alpha(j_{1},j_{2},j_{3},j_{4}+1,j_{5},j_{6})\,,
\ee
\begin{multline}
\alpha
\,=\,
\f{1}{j_{4}^{2}}
\sqrt{(j_{4}+j_{5}-j_{3})(j_{4}-j_{5}+j_{3})(1-j_{4}+j_{5}+j_{3})(1+j_{4}+j_{5}+j_{3})}
\\
\times
\sqrt{(j_{4}+j_{2}-j_{6})(j_{4}-j_{2}+j_{6})(1-j_{4}+j_{2}+j_{6})(1+j_{4}+j_{2}+j_{6})}\,,
\end{multline}
\be
\alpha_{0}
\,=\,
-\f{\Big{[}j_{4}(j_{4}+1)+j_{5}(j_{5}+1)-j_{3}(j_{3}+1)\Big{]}\Big{[}j_{4}(j_{4}+1)+j_{2}(j_{2}+1)-j_{6}(j_{6}+1)\Big{]}}{4j_{4}(j_{4}+1)}\,.
\ee
So we would like to compare the action of the triple grasping to the classical volume:
\be
\f1{3!}\hat{T}_{456}\sixj{j_{1}}{j_{2}}{j_{3}}{j_{4}}{j_{5}}{j_{6}}
\quad
\overset{??}{\longleftrightarrow}
\quad
\sqrt{U}\,\sixj{j_{1}}{j_{2}}{j_{3}}{j_{4}}{j_{5}}{j_{6}}\,.
\ee
Plotting the ratio between the triple grasping and the tetrahedron volume, one realizes that they do not match, even for large spins. This was explained in \cite{Hackett:2006gp} due to the fact that the triple grasping acts as a odd order differential operator shifting the $\cos(S_{R}+\f\pi 4)$ oscillatory asymptotics of the 6j-symbol to a off-phase $\sin(S_{R}+\f\pi 4)$ factor (where $S_{R}$ is the Regge action for the tetrahedron).
This can be remedied by considering an even power of the grasping operator allows to restore the correct oscillatory behavior. So we compare the squared triple grasping to the classical squared volume:
\be
\f1{3!^{2}}\hat{T}_{456}^{2}\sixj{j_{1}}{j_{2}}{j_{3}}{j_{4}}{j_{5}}{j_{6}}
\quad
\overset{??}{\longleftrightarrow}
\quad
U\,\sixj{j_{1}}{j_{2}}{j_{3}}{j_{4}}{j_{5}}{j_{6}}\,.
\ee
As pointed out in \cite{Hackett:2006gp}, the squared triple grasping has an ordering issue and the natural choice is to make the second graspings act without interfering with the first ones, thus taking the following convention:
\begin{multline}
\hat{T}_{456}^{2}\sixj{j_{1}}{j_{2}}{j_{3}}{j_{4}}{j_{5}}{j_{6}}
=
\alpha_{-}(j_{4})\alpha_{-}(j_{4}-1)\sixj{j_{1}}{j_{2}}{j_{3}}{j_{4}-2}{j_{5}}{j_{6}}
+\alpha_{+}(j_{4})\alpha_{+}(j_{4}+1)\sixj{j_{1}}{j_{2}}{j_{3}}{j_{4}+2}{j_{5}}{j_{6}}
\\
+\big{(}\alpha_{0}^{2}+\alpha_{-}(j_{4})\alpha_{+}(j_{4}-1)+\alpha_{-}(j_{4}+1)\alpha_{+}(j_{4})\big{)}
\sixj{j_{1}}{j_{2}}{j_{3}}{j_{4}}{j_{5}}{j_{6}}
\,.
\end{multline}
A plot of the ratio between the squared triple grasping (rescaled by $3!^2$) and the squared volume is given in fig.\ref{fig:triplegrasping}: the squared triple grasping gives minus the squared volume at large spins  as expected. We see nevertheless large oscillating deviations coming from probably almost-zeroes of the 6j-symbol. It does not seem possible to get rid of these oscillations in a simple way. We also plot in fig.\ref{fig:shiftedvolume} the same ratio but taking the volume of the shifted tetrahedron with edge lengths $j_{i}+\f12$ (as expected from the exact asymptotic formula for the 6j-symbol). This clearly improves the convergence of the ratio to 1.


\bibliographystyle{bib-style}
\bibliography{qDeformation}

\end{document}